\documentclass{book}
\usepackage{makeidx,epsfig}
\usepackage{setspace,graphicx}
\usepackage{Generic}

\makeindex

\usepackage{natbib,amsmath,bbm,amssymb}
\usepackage{amsthm,amscd,multirow,subfigure,graphics}
\usepackage[labelsep=period]{caption}

\def\OLH{{\mbox{OLH}}}

\def\D{{\mbox{D}}}

\def\OA{{\mbox{OA}}}

\newtheorem{theorem}{Theorem}[section]

\newtheorem{corollary}{Corollary}[section]

\newtheorem{example}{Example}[section]
\newcommand{\bX}{\mbox{\boldmath{$X$}}}
\newcommand{\bx}{\mbox{\boldmath{$x$}}}

\newcommand{\ba}{\mbox{\boldmath{$a$}}}

\newcommand{\bD}{\mbox{\boldmath{$D$}}}
\newcommand{\bd}{\mbox{\boldmath{$d$}}}
\newcommand{\bA}{\mbox{\boldmath{$A$}}}
\newcommand{\bB}{\mbox{\boldmath{$B$}}}
\newcommand{\bE}{\mbox{\boldmath{$E$}}}
\newcommand{\bF}{\mbox{\boldmath{$F$}}}
\newcommand{\bu}{\mbox{\boldmath{$u$}}}
\newcommand{\bh}{\mbox{\boldmath{$h$}}}
\newcommand{\bv}{\mbox{\boldmath{$v$}}}

\newcommand{\bL}{\mbox{\boldmath{$L$}}}

\newcommand{\bV}{\mbox{\boldmath{$V$}}}
\newcommand{\bI}{\mbox{\boldmath{$I$}}}
\newcommand{\bS}{\mbox{\boldmath{$S$}}}
\newcommand{\bT}{\mbox{\boldmath{$T$}}}
\newcommand{\bH}{\mbox{\boldmath{$H$}}}
\newcommand{\bR}{\mbox{\boldmath{$R$}}}
\newcommand{\bU}{\mbox{\boldmath{$U$}}}

\newcommand{\bJ}{\mbox{\boldmath{$J$}}}
\newcommand{\bzero}{\textbf{0}}

\newcommand{\bchi}{\mbox{ \boldmath${\chi}$} }
\newenvironment{myindentpar}[1]%
{\begin{list}{}%
         {\setlength{\leftmargin}{#1}}%
         \item[]%
}
{\end{list}}

\bibpunct{(}{)}{;}{a}{}{,}

\begin{document}
\title{Latin Hypercubes and Space-filling Designs}
\author{C.\ Devon Lin  and Boxin Tang}
\date{Feb 16, 2014}
\pagenumbering{roman}
\begin{doublespace}
\maketitle
\frontmatter
\tableofcontents
\mainmatter
\pagenumbering{arabic}
\setcounter{chapter}{19}
\maketitle
\markboth{}{{Chapter 19 \ Latin Hypercubes and Space-filling Designs}}

\section{Introduction}\label{sec:intro}

This chapter discusses a general design approach to planning computer experiments, which seeks   design points  that fill a bounded   design region as uniformly as possible. Such designs are broadly referred to as {\em space-filling designs}.

The literature on the design for computer experiments has focused mainly on deterministic computer models; that is, running computer code with the same inputs always produces the same outputs (see Chapter 17).
Because of this feature, the three fundamental design principles, {\em randomization}, {\em replication} and {\em blocking}, are irrelevant in computer experiments.  The true relationship between the inputs and the responses is unknown and often very complicated.
To explore the relationship, one could use traditional regression models. But the most popular are Gaussian process models; see Chapter 17 for details. However, before data are collected,  quite often little {\em a priori} or background knowledge is available about which model would be appropriate, and designs for computer experiments should facilitate diverse modelling methods. For this purpose, a space-filling design is the best choice. The design region in which to make prediction may be unspecified at the data collection stage. Therefore, it is   appropriate to use designs that represent all portions of the design region.  When the primary goal of experiments is to make prediction at unsampled points, space-filling designs allow us to build a predictor with better average accuracy.

One most commonly used class of space-filling designs for computer experiments is  that of  Latin hypercube designs. Such designs, introduced by \cite{mckay1979comparison},  do not have repeated runs.  Latin hypercube designs have one-dimensional uniformity in that, for each input variable, if its range is divided into the same number of
equally-spaced intervals as the number of observations, there is exactly one observation in each interval.
However, a random Latin hypercube design may not be a good choice with respect to some optimality criteria such as maximin distance and orthogonality (discussed later).
The maximin distance criterion, introduced by \cite{johnson1990minimax}, maximizes the smallest distance between any two design points so that no two design points are too close. Therefore, a maximin distance design  spreads out its points evenly over the entire design region. To further enhance the space-filling property for each individual input of a maximin distance design, \cite{morris1995exploratory} proposed the use of maximin Latin hypercube designs.

Many applications involve a large number of input variables.
Finding space-filling designs with a limited number of design points
that provide a good coverage of the entire high dimensional input space is a very ambitious, if not hopeless, undertaking.  A more reasonable approach is to construct designs that are space-filling in the low dimensional projections.  \cite{MoonDeanSantner2011} constructed designs that are space-filling in the two-dimensional projections and demonstrated empirically that such designs also perform well in terms of the maximin distance criterion in higher dimensions. Other designs that are space-filling in the low dimensional projections are
randomized orthogonal arrays \cite[]{owen1992central} and orthogonal array-based Latin hypercubes \cite[]{tang1993orthogonal}.  Another important approach is to construct orthogonal Latin hypercube designs.  The basic idea of this approach is that orthogonality can be viewed as a stepping stone to constructing designs that are space-filling in low dimensional projections \cite[]{bingham2009orthogonal}.

Originating as  popular tools in numerical analysis, low-discrepancy nets, low-discrepancy sequences and uniform designs have also been  well recognized as space-filling designs for computer experiments. These designs are chosen to achieve  uniformity in the design space based on the discrepancy criteria such as the $L_p$ discrepancy (see Section~\ref{subsec:other2}).

As an alternative to the use of space-filling designs, one could choose designs that perform well with respect to some model-dependent criteria such as the minimum integrated mean square error and the maximum entropy \cite[]{sacks1989design,shewry1987maximum}. One drawback of this approach is that such designs require the prior knowledge of the model. For instance, to be able to construct maximum entropy designs and  integrated mean square error  optimal designs, one would need the values of
the parameters in the correlation function when a Gaussian process is used to model responses. One could also consider a Bayesian approach \cite[]{Leatherman2014tr}. A detailed account of model-dependent designs can be found in \cite{santner2003design}, \cite{fang2006design} and the references therein.

This chapter is organized as follows. Section~\ref{sec:LHD} gives a detailed review of Latin hypercube designs, and discusses three important types of Latin hypercube designs (Latin hypercube designs based on measures of distance; orthogonal array-based Latin hypercube designs; orthogonal and nearly orthogonal Latin hypercube designs).  Section~\ref{sec:other} describes other space-filling designs that are not Latin hypercube designs.
Concluding remarks are provided in Section~\ref{sec:con}.

\section{Latin Hypercube Designs}\label{sec:LHD}

\subsection{Introduction and examples}\label{subsec:intro}

\begin{figure}[!htb]
\centering
\includegraphics[width=0.7\textwidth]{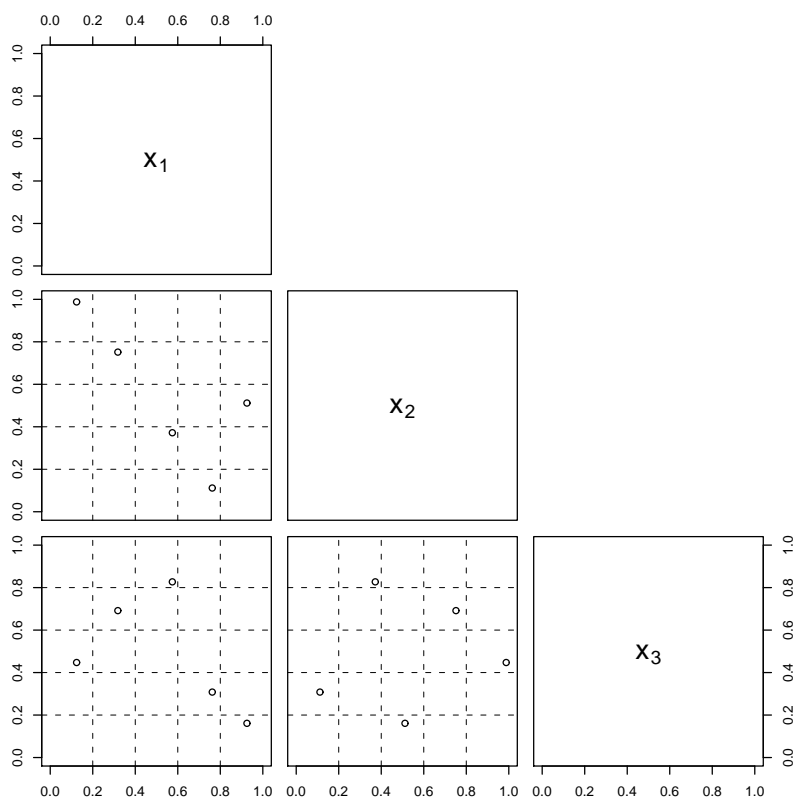}
\caption{The pairwise plot of the Latin hypercube design $\bD$ in Table~\ref{tab:lh1} for the three factors $x_1,x_2,x_3$}\label{fig:lhs1}
\end{figure}

A Latin hypercube of $n$ runs for $k$ factors is represented by an $n \times k$ matrix, each column of which is a permutation
of $n$ equally spaced levels. For convenience, the $n$ levels are taken to be $-(n-1)/2, -(n-3)/2, \ldots, (n-3)/2,  (n-1)/2$. For example,  design $\bL$ in Table~\ref{tab:lh1} is a Latin hypercube of $5$ runs for $3$ factors. Given an $n \times k$ Latin hypercube  $\bL=(l_{ij})$, a Latin hypercube design $\bD$ in the design space $[0,1)^k$ can be generated and the design matrix of $\bD$ is an $n \times k$ matrix with the $(i,j)$th entry being
\begin{eqnarray}\label{eq:dij}
d_{ij} = \frac{ l_{ij} + (n-1)/2 + u_{ij} }{n}, \ \ \ i=1, \ldots, n, j = 1, \ldots, k,
\end{eqnarray}
\noindent where $u_{ij}$'s are independent random numbers from $[0,1)$.
If each $u_{ij}$ in (\ref{eq:dij}) is taken to be 0.5, the resulting design $\bD$ is termed ``lattice sample'' due to \cite{patterson1954errors}.
For each factor, Latin hypercube designs have exactly one point  in each of the $n$  intervals $[0,1/n), [1/n,2/n),\ldots,[(n-1)/n,1)$. This property is referred to as {\em one-dimensional uniformity}.  For instance, design $\bD$  in Table~\ref{tab:lh1} is a Latin hypercube design based on the $\bL$ in the table, and its pairwise plot in Figure~\ref{fig:lhs1}  illustrates the
one-dimensional uniformity. When the five points are projected onto each axis, there is exactly one point in each of the five equally-spaced intervals.

\begin{table}[!htb]
\begin{center}
\caption{A $5 \times 3$ Latin hypercube $\bL$   and a Latin hypercube
design $\bD$ based on $\bL$}
\begin{tabular}{rrrrrrrr}
\multicolumn{3}{c}{$\bL$} & \multicolumn{2}{c}{} & \multicolumn{3}{c}{$\bD$}\\
   2&    0 &   -2 & &  $\quad \quad $ &  0.9253& 0.5117&0.1610 \\
   1&   -2 &   -1 & & $\quad \quad $ &  0.7621& 0.1117&0.3081 \\
  -2&    2 &    0 & & $\quad \quad $ &  0.1241& 0.9878&0.4473 \\
   0&   -1 &    2 & & $\quad \quad $ &  0.5744& 0.3719&0.8270 \\
  -1&    1 &    1 & & $\quad \quad $ &  0.3181& 0.7514&0.6916 \\
\end{tabular}\label{tab:lh1}
\end{center}
\end{table}

The popularity of Latin hypercube designs was largely attributed to their theoretical justification for the variance reduction in numerical integration.  Consider a function $y=f(\bx)$ where $f$ is known, $\bx=(x_1,\ldots,x_k)$ has a uniform distribution in the unit hypercube $[0,1)^k$, and
$y \in R$. (More generally, when $x_j$ follows a continuous distribution with a cumulative distribution function $F_j$, then the inputs of $x_j$ can be selected via the quantile transformation $F_j^{-1}(u_j)$ where $u_j$ follows a uniform distribution in $[0,1)$).
The expectation of $y$,
\begin{eqnarray}\label{eq:mu}
\mu=\hbox{E}(y),
\end{eqnarray}
\noindent is of interest. When the expectation $\mu$ cannot be computed explicitly or its derivation is unwieldy, one can resort to approximate methods. Let $\bx_1,\ldots,\bx_n$ be a sample of size $n$. One estimate of $\mu$ in (\ref{eq:mu}) is
\begin{eqnarray}\label{eq:muhat}
\hat{\mu} = \frac{1}{n} \sum_{i=1}^n f(\bx_i).
\end{eqnarray}
The approach that takes $\bx_1,\ldots,\bx_n$   independently from the uniform distribution in $[0,1)^k$ is {\em simple random sampling}.
\cite{mckay1979comparison} suggested an approach based on a Latin hypercube  sample $\bx_1, \ldots,\bx_n$.
Denote the estimator $\hat{\mu}$ in  (\ref{eq:muhat})  of $\mu$ under  simple random sampling and  Latin hypercube sampling by $\hat{\mu}_{srs}$ and $\hat{\mu}_{lhs}$, respectively. Note that $\hat{\mu}_{srs}$ and $\hat{\mu}_{lhs}$ have the same form but $\hat{\mu}_{srs}$ uses   a simple random sample and $\hat{\mu}_{lhs}$  a Latin hypercube sample. Both samples are denoted by $\bx_1, \ldots, \bx_n$, for convenience.
\cite{mckay1979comparison} established the following theorem.
\begin{theorem}\label{theo:var}
If $y=f(\bx)$ is monotonic in each of its input variables, then $\hbox{Var}(\hat{\mu}_{lhs} ) \leq \hbox{Var}(\hat{\mu}_{srs}).$
\end{theorem}

Theorem~\ref{theo:var} says that when the monotonicity condition holds, Latin hypercube sampling yields a smaller variance of the sample mean than  simple random sampling.  Theorem~\ref{theo:var2} below \cite[]{stein1987large} provides
  some insights into the two methods of sampling.
\begin{theorem}\label{theo:var2}
We have that for $\bx \in [0,1)^k$,
$$ \hbox{Var}(\hat{\mu}_{srs}) = \frac{1}{n}\hbox{Var}[f(\bx)]$$
and
$$\hbox{Var}(\hat{\mu}_{lhs}) = \frac{1}{n} \hbox{Var}[f(\bx)] - \frac{1}{n} \sum_{j=1}^k \hbox{Var}[f_j(x_j)] +o(\frac{1}{n}),$$
\noindent where $x_j$ is the $j$th input of $\bx$, $f_j(x_j) = \hbox{E}[f(\bx)|x_j]-\mu$ and $o(\cdot)$ is little o notation.
\end{theorem}

The term $f_j(x_j)$ in Theorem~\ref{theo:var2} is  the main effect of the $j$th input variable.
 Theorem~\ref{theo:var2} tells us that the variance of the sample mean under Latin hypercube sampling is smaller than
 the counterpart under simple random sampling by an amount contributed by main effects. The extent of the variance reduction depends on the extent to which the function $f$ is additive in the inputs. Asymptotic normality and a central limit theorem of Latin hypercube sampling were established in \cite{stein1987large} and \cite{owen1992central}, respectively. A related approach is that of quasi-Monte Carlo methods, which selects design points in a deterministic fashion \cite[see][and Section 19.3.2]{Niederreiter1992}.

A randomly generated Latin hypercube design does not necessarily perform well with respect to criteria such as those of ``space-filling'' or ``orthogonality'', alluded to in Section~\ref{sec:intro}.  For example, when projected onto two factors, design points in a random Latin hypercube design may roughly lie on the diagonal as in the plot of $x_1$ versus $x_2$ in Figure~\ref{fig:lhs1}, leaving a large area in the design space unexplored. In this case, the corresponding two columns in the design matrix are  highly correlated.  Examples of Latin hypercube designs with desirable properties are maximin Latin hypercube designs, orthogonal-array based Latin hypercube designs, and orthogonal or nearly orthogonal Latin hypercube designs; these will be discussed throughout the chapter.

\subsection{Latin hypercube designs based on measures of distance}\label{subsec:maximin}

To construct space-filling Latin hypercube designs, one natural approach is to make use of distance criteria. In what follows, we review several measures of distance.

Let $\bu=(u_1,\ldots,u_k)$ and $\bv=(v_1,\ldots,v_k)$ be  two design points in the design space $\bchi=[0,1]^k$ . For $t>0$, define the inter-point distance between $\bu$ and $\bv$
to be
\begin{equation}\label{eq:d}
d(\bu,\bv)= \big (\sum_{j=1}^k|u_j-v_j|^t\big )^{1/t}.
\end{equation}
\noindent When $t=1$ and $t=2$, the measure in (\ref{eq:d}) becomes the rectangular and Euclidean distances, respectively.
The {\em maximin distance} criterion seeks  a design $\bD$ of $n$ points in the design space $\bchi$ that maximizes the smallest inter-point distance; that is, it maximizes
\begin{equation}\label{eq:maximin}
\min_{\substack{\bu, \bv \in \bD \\ {\bu \neq \bv} }} d(\bu,\bv),
\end{equation}
\noindent where $d(\bu,\bv)$ is defined as in (\ref{eq:d}) for any given $t$. This criterion attempts to place the design points such that  no two  points are too close to each other.

A slightly different idea is to spread out the design points of a design $\bD$ in such a way that every point  in the design space $\bchi$ is close to some point in $\bD$.  This is  the {\em minimax distance} criterion which seeks a design $\bD$ of $n$ points in $\bchi$ that minimizes the maximum distance between an arbitrary point $\bx \in \bchi$ and the design $\bD$; that is, it minimizes $$ \max_{\bx \in \bchi} d(\bx,\bD),$$
\noindent where $d(\bx, \bD)$, representing the distance between $\bx$ and the closest point in $\bD$, is defined as $d(\bx, \bD) = \min_{\bx_i \in \bD} d(\bx,\bx_i)$ and $d(\bx,\bx_i)$ is given in (\ref{eq:d}) for any given $t$.

\cite{audze1977new} introduced a distance criterion  similar in spirit to the maximin distance criterion by using
\begin{eqnarray}\label{eq:aefunc}
\sum_{1 \leq i <j \leq n} d(\bx_i,\bx_j)^{-2},
\end{eqnarray}
\noindent where $\bx_1,\ldots, \bx_n$ are the design points. This criterion of minimizing (\ref{eq:aefunc}) was used by \cite{liefvendahl2006study}.

\cite{MoonDeanSantner2011} defined a two-dimensional maximin distance criterion.
Let the inter-point distance between two design points $\bu=(u_1,\ldots,u_k)$ and $\bv=(v_1,\ldots,v_k)$ projected onto dimensions $h$ and $l$
 be
\begin{equation*}
d^{(2)}_{h,l}(\bu,\bv)= \big (|u_h-v_h|^t+|u_l-v_l|^t\big )^{1/t}, \ \ t>0.
\end{equation*}
Then the minimum inter-point distance of a design $\bD$ over all two-dimensional subspaces is
\begin{equation}\label{eq:dmin2}
d_{min}^{(2)} = \min_{\substack{\bu, \bv \in \bD \\ {\bu \neq \bv},h \neq l}} d^{(2)}_{h,l}(\bu,\bv).
\end{equation}
\noindent The two-dimensional maximin distance criterion selects a design that maximizes $d_{min}^{(2)}$ in (\ref{eq:dmin2}). \cite{MoonDeanSantner2011} showed by examples that optimal Latin hypercube designs based on this criterion also perform well under the maximin distance criterion (\ref{eq:maximin}).

\subsubsection{Maximin Latin hypercube designs}\label{subsubsec:maximin}
We now focus on maximin distance criterion.
 Recall the Gaussian process model in Section 17.4.1,
\begin{eqnarray}\label{eq:ok}
Y(x) = \mu + Z(x),
\end{eqnarray}
\noindent where $\mu$ is the unknown but constant mean function, $Z(x)$ is a stationary Gaussian process with mean 0, variance $\sigma^2$, and correlation function $R(\cdot | \theta)$. A popular choice for the correlation function is the power exponential correlation
\begin{eqnarray*}\label{eq:cor}
 R(\bh| \theta) = \hbox{exp}\big  ( - \theta \sum_{j=1}^k  |h_j|^p\big ),  \ \  0 < p \leq 2,
\end{eqnarray*}
\noindent where $h_j$ is the $j$th element of $\bh$. \cite{johnson1990minimax} showed that as the correlation parameter $\theta$ goes to infinity,  a maximin design maximizes the determinant of the correlation matrix, where the correlation matrix refers to that of the outputs from running the computer model at the design points. That is, a maximin design is asymptotically D-optimal under the model in (\ref{eq:ok}) as the correlations become weak. Thus, a maximin design is also asymptotically optimal with respect to the maximum entropy criterion \cite[]{shewry1987maximum}.

The problem of finding maximin designs is referred to as the maximum facility dispersion problem \cite[]{erkut1990discrete} in location theory. It is closely related to
the sphere packing problem in the field of discrete and computational geometry \cite[]{melisse1997,conway1999sphere}.
The two problems are, however, different as explained in
\cite{johnson1990minimax}.

An extended definition of a maximin design was given by \cite{morris1995exploratory}. Define a distance list $(d_1,\ldots,d_m)$ and  an index list $(J_1,\ldots,J_m)$ respectively in the following way. The distance list contains the distinct values of inter-point distances, sorted from the smallest to the largest, and $J_i$ in the index list is the number of pairs of design points in the design separated by the  distance $d_i$, $i=1,\ldots, m$. Note that $1 \leq m \leq  {n \choose 2}$. In \cite{morris1995exploratory}, a design is called a maximin design if it sequentially maximizes $d_i$'s and minimizes $J_i$'s in the order $d_1,J_1,d_2,J_2,\ldots,d_m,J_m$. They further introduced a computationally efficient scalar-value criterion
\begin{eqnarray}\label{eq:phi}
\phi_q=\big (\sum_{i=1}^m\frac{J_i}{d_i^{q}}\big)^{1/q},
\end{eqnarray}
\noindent where $q$ is a positive integer. Minimizing $\phi_q$ with a large $q$ results in a maximin design. Values of $q$ are chosen  depending on the size of the design searched for, ranging from 5 for small designs to 20 for moderate-sized designs to 50 for large designs.
\begin{figure}[!htb]
\centering
\subfigure[]{\label{fig:a} \includegraphics[width=0.45\textwidth]{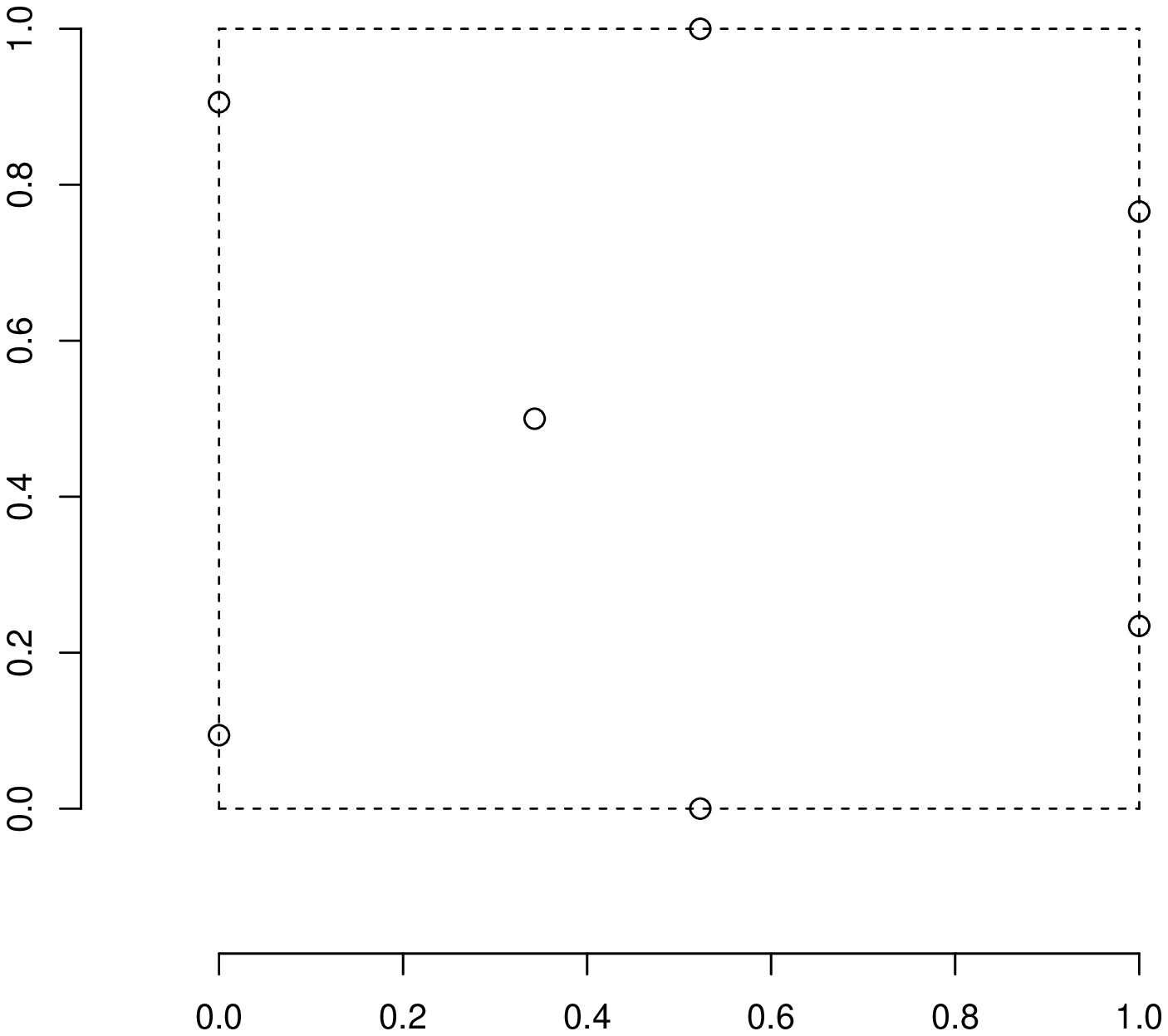} }\hspace{0.1in}
\subfigure[]{\label{fig:b}  \includegraphics[width=0.45\textwidth]{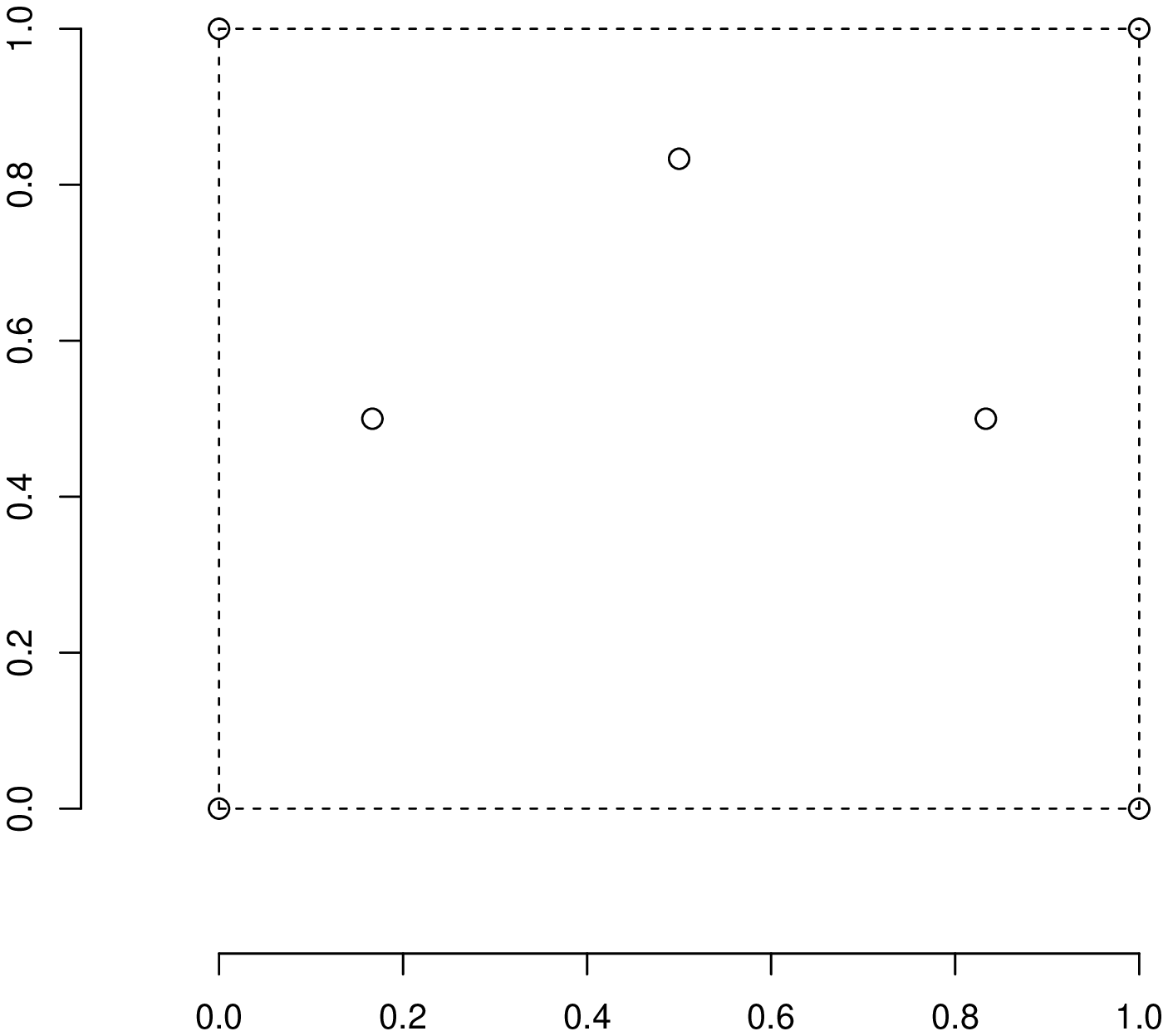}}
\caption{Maximin designs with $n=7$ points on $[0,1]^2$: (a) Euclidean distance;
 (b) rectangular distance}\label{fig:maximin}
\end{figure}
\begin{table}[!htb]
\begin{center}
\caption{Algorithms for generating maximin Latin hypercube designs}
\scalebox{0.82}{
\begin{tabular}{l l l l l}
\hline\\
Article & $\quad$  &  Algorithm &  $\quad$  &Criterion\\[2pt]
\hline
\cite{morris1995exploratory} & & simulated annealing && $\phi_q^{(a)}$ \\[2pt]
\cite{ye2000algorithmic} & & columnwise-pairwise &  &$\phi_q$ and entropy \\[2pt]
\cite{jin2005efficient} & & enhanced stochastic  &  & $\phi_q$, entropy and $L_2$ discrepancy \\[2pt]
 & & \ \ \ evolutionary algorithm & & \\[2pt]
\cite{liefvendahl2006study} & & columnwise-pairwise and & & maximin and the Audze-Eglais \\[2pt]
& &  \ \ \  genetic algorithm & &  \ \ \  function$^{(b)}$\\[2pt]
\cite{van2007maximin} &&  branch-and-bound & & maximin with Euclidean distance\\[2pt]
\cite{Forrester} && evolutionary operation & &$\phi_q$ \\[2pt]
\cite{grosso2009finding} & &iterated local search heuristics & & $\phi_q$ \\[2pt]
\cite{viana2010algorithm} && translational propagation & & $\phi_q$ \\[2pt]
\cite{zhu2011novel} && successive local enumeration && maximin\\[2pt]
\cite{MoonDeanSantner2011} && smart swap algorithm && ${d_{min}^{(2)}}^{(c)}$, $\phi_q$\\[2pt]
\cite{chen2012optimizing} && particle swarm algorithm &&  $\phi_q$\\[2pt]
\hline
\multicolumn{5}{l}{\small (a): $\phi_q$ is given as in (\ref{eq:phi}); (b): the Audze-Eglais function in (\ref{eq:aefunc}); (c) $d_{min}^{(2)}$ is given in (\ref{eq:dmin2}).}
\end{tabular}}\label{tab:alg}
\end{center}
\end{table}

Maximin designs tend to place design points toward or on the boundary. For example, Figure~\ref{fig:maximin} exhibits a maximin Euclidean distance design and
a maximin rectangular distance design, both of seven  points.
Maximin designs are likely to have clumped projections onto one-dimension.  Thus, such designs may not possess desirable one-dimensional uniformity which is guaranteed by Latin hypercube designs. To strike the balance,  \cite{morris1995exploratory} examined maximin designs within Latin hypercube designs. Although this idea sounds simple,
generating maximin Latin hypercube designs is a challenging task particularly for large designs.  The primary approach for obtaining maximin Latin hypercube designs is using the algorithms  summarized in Table~\ref{tab:alg}. These algorithms search for maximin Latin hypercube designs that have $u_{ij}$ in (\ref{eq:dij}) being a constant. For example, \cite{MoonDeanSantner2011} used $u_{ij}=0.5$, which corresponds to the midpoint of the interval $[(i-1)/n,i/n]$ for $i=1,\ldots,n$.
For detailed descriptions of these algorithms, see the respective references.

Some available implementations of the algorithms in Table~\ref{tab:alg} include
the Matlab code provided in \cite{viana2010algorithm}, the function {\em maximinLHS} in the R package {\em lhs} \cite[]{carnell2009lhs}, and the function {\em lhsdesign} in the Matlab statistics toolbox.  The function in R uses 
random $u_{ij}$'s in (\ref{eq:dij}) while the function in Matlab allows both random $u_{ij}$'s and $u_{ij}=0.5$.
It should be noted, however, that these designs are approximate maximin Latin hypercube designs. No theoretical method is available to construct exact maximin Latin hypercube designs of flexible run sizes except  \cite{tang1994theorem} and \cite{van2007maximin}. These two articles provided methods for constructing exact  maximin  Latin hypercube designs of certain run sizes and  numbers of input variables. \cite{tang1994theorem} constructed a Latin hypercube based on  a single replicate  full factorial design \cite[see Chapter 1 and also][Chapter 4]{wu2011experiments} and showed that the constructed Latin hypercube has the same rectangular distance as the single replicate full factorial design, where the rectangular distance of a design is given by (\ref{eq:maximin}) with $t=1$.
\cite{van2007maximin} constructed two-dimensional maximin Latin hypercubes with the distance measures with $t=1$ and $t= \infty$ in (\ref{eq:d}). For the Euclidean distance measure with $t=2$,  \cite{van2007maximin} used the branch-and-bound algorithm to find maximin Latin hypercube designs with $n\leq 70$. 

\subsection{Orthogonal array-based Latin hypercube designs}   \label{subsec:OALHD}

\cite{tang1993orthogonal} proposed orthogonal array-based Latin hypercube designs, also known as U designs, which guarantee multi-dimensional space-filling. Recall the definition of an $s$-level orthogonal array (OA) of $n$ runs, $k$ factors and strength $r$, denoted by $\OA(n,s^k,r)$  in Chapter~9.
The $s$ levels are taken to be $1,2,\ldots,s$ in this chapter.
By the definition of orthogonal arrays, a Latin hypercube of $n$ runs for $k$ factors is an $\OA(n,n^k,1)$.
\begin{table}[!htb]
\begin{center}
\caption{An $\OA(9,3^4,2)$ and a corresponding OA-based Latin hypercube}
\begin{tabular}{rrrr rr rrrr}
\multicolumn{4}{c}{ $\OA(9,3^4,2)$}&  $\quad$  & $\quad$ & \multicolumn{4}{c}{ $\bL$}\\
1 &1 &1 &1 &  $\quad$  & $\quad$ & -2 &  -2  & -4 &  -2    \\
1 &2 &2 &3   &  $\quad$  & $\quad$ &    -4 &   0  &  1 &   2    \\
1 &3 &3 &2 &  $\quad$  & $\quad$ & -3 &   4  &  2 &   1    \\
2 &1 &2 &2&  $\quad$  & $\quad$ &  -1 &  -4  & -1 &  -1    \\
2 &2 &3 &1&  $\quad$  & $\quad$ &    1 &  -1  &  4 &  -3    \\
2 &3 &1 &3 &  $\quad$  & $\quad$ &   0 &   2  & -3 &   4    \\
3 &1 &3 &3 &  $\quad$  & $\quad$ &      3 &  -3  &  3 &   3    \\
3 &2 &1 & 2  &  $\quad$  & $\quad$ &    2 &   1  & -2 &   0    \\
3 &3 &2 &1 &  $\quad$  & $\quad$ &     4 &   3  &  0 &  -4    \\
\end{tabular}\label{tab:oa1}
\end{center}
\end{table}

The construction of OA-based Latin hypercubes in \cite{tang1993orthogonal} works as follows. Let $\bA$ be an $\OA(n,s^k,r)$. For each column of $\bA$ and $m=1,\ldots,s$, replace the $n/s$ positions with entry $m$ by a random permutation of $(m-1)n/s+1, (m-1)n/s+2,\ldots, mn/s$.
Denote the design after the above replacement procedure by $\bA'$. In our notation,  an OA-based Latin hypercube is given by $\bL = \bA' - (n+1)\bJ/2$, where $\bJ$ is an $n\times k$ matrix of all 1's. An OA-based Latin hypercube design in the design space $[0,1)^k$ can be generated via (\ref{eq:dij}). In addition to the one-dimensional uniformity, an $\OA(n,s^k,r)$-based Latin hypercube has the $r$-dimensional projection property that when projected onto any $r$ columns, it has exactly  $\lambda=n/s^r$ points in each of the  $s^r$ cells $\mathcal{P}^r$ where $\mathcal{P}=\{[0,1/s],[1/s,2/s), \ldots, [1-1/s,1)\}$.
Example~\ref{exam:oalhs} illustrates this feature of an $\OA(9,3^4,2)$-based Latin hypercube.
\begin{example}\label{exam:oalhs}
Table~\ref{tab:oa1} displays an OA-based Latin hypercube $\bL$ based on the orthogonal array $\OA(9,3^4,2)$ in the table.  Figure~\ref{fig:oalhs1} depicts the pairwise plot of this  Latin hypercube. In each subplot, there is exactly one point in each of nine dot-dash line boxes.
\begin{figure}[!htb]
\centering
\includegraphics[width=0.65\textwidth]{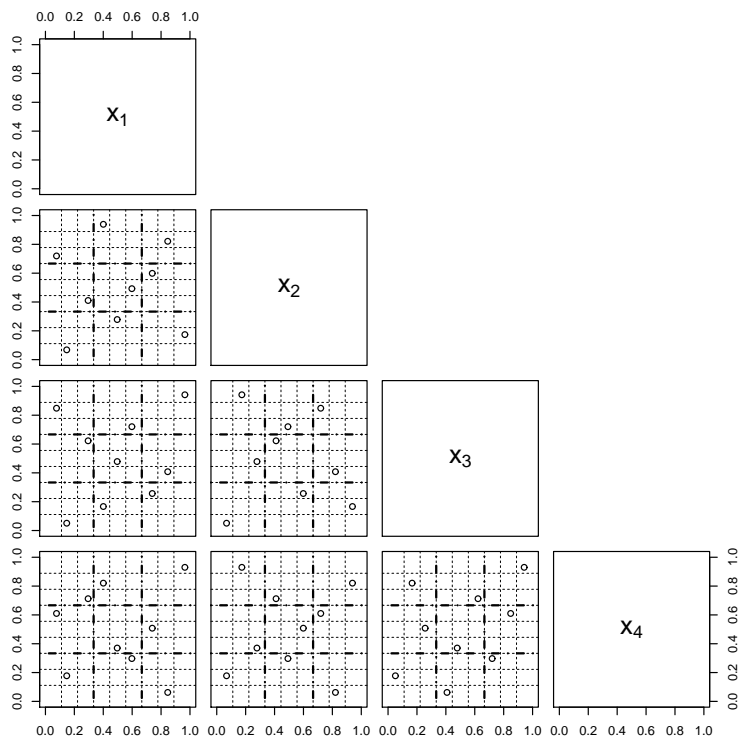}
\caption{The pairwise plot of an OA-based Latin hypercube design based on the Latin hypercube in Table~\ref{tab:oa1} for the four factors $x_1,\ldots, x_4$ }\label{fig:oalhs1}
\end{figure}
\end{example}

\begin{figure}[!htb]
\centering
\subfigure[]{\label{fig:lhsa} \includegraphics[width=0.4\textwidth]{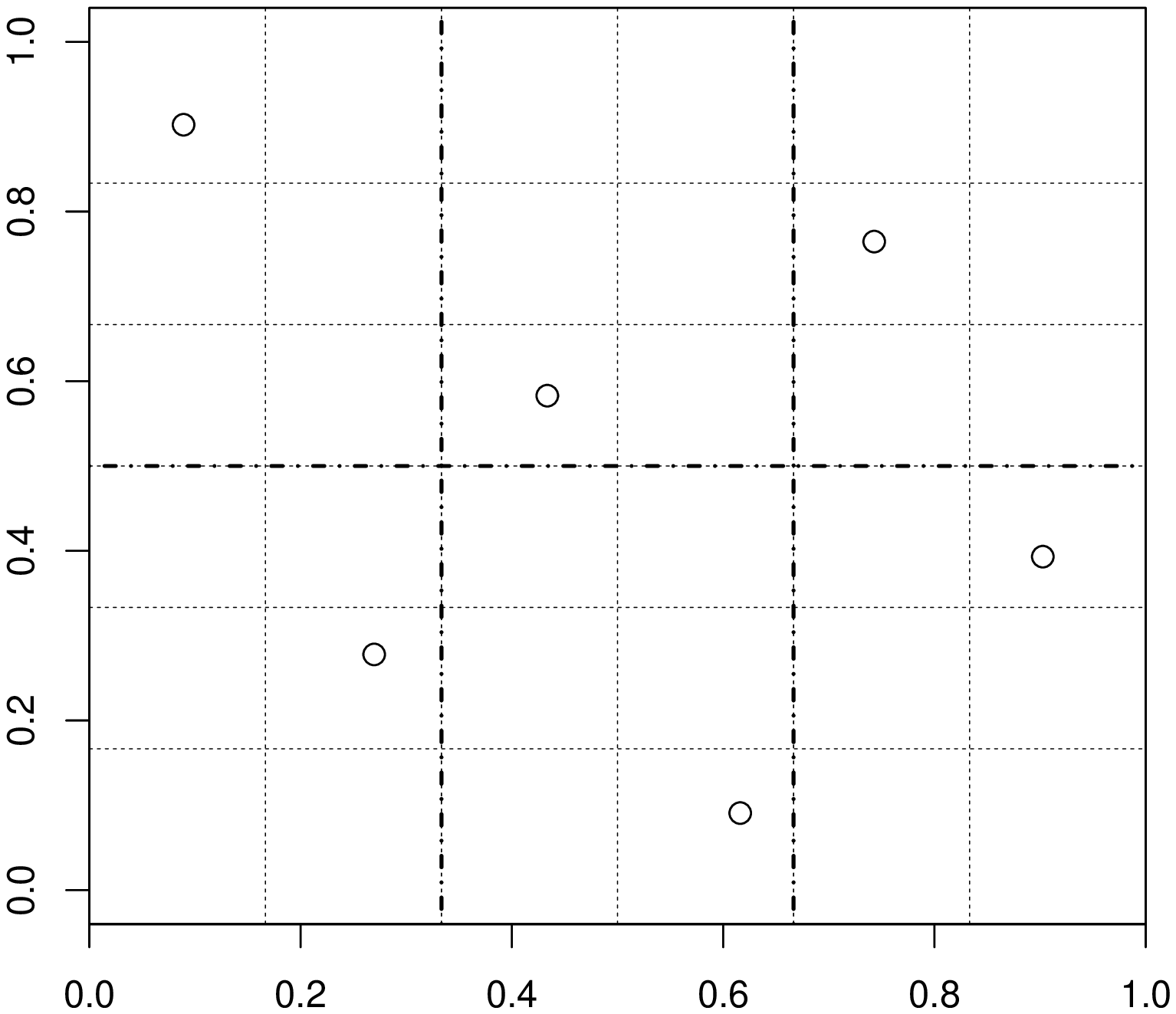} }\hspace{0.1in}
\subfigure[]{\label{fig:lhsb}  \includegraphics[width=0.4\textwidth]{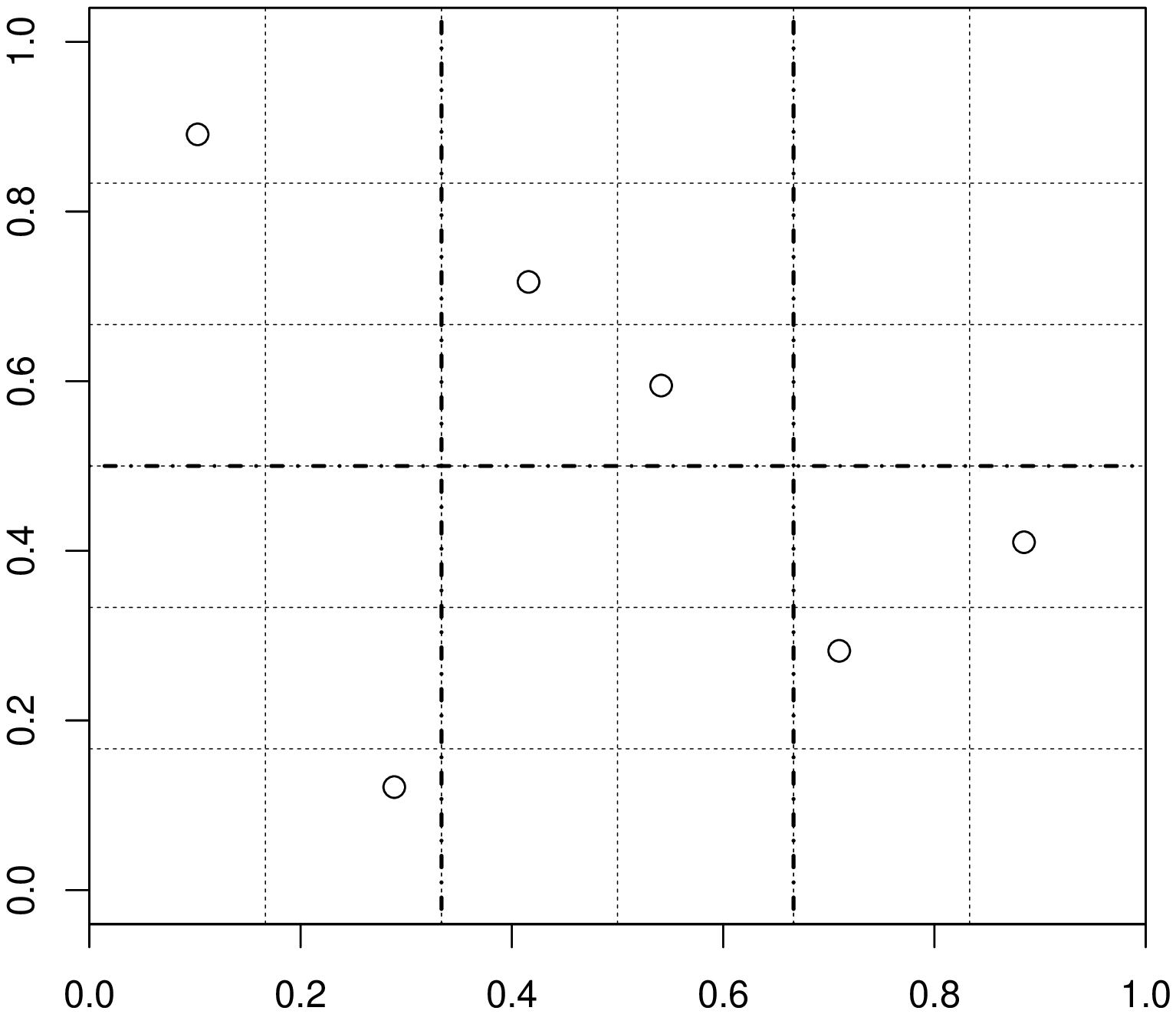}}
\caption{Plots of six-point Latin hypercube designs: (a) a  Latin hypercube design based on an asymmetrical orthogonal array; (b)  a random Latin hypercube design }\label{fig:oalhs2}
\end{figure}

A generalization of OA-based Latin hypercubes using asymmetrical orthogonal arrays (see Chapter 9) can be readily made.
For example, Figure~\ref{fig:lhsa}  displays a Latin hypercube design based on an asymmetrical orthogonal array of six runs for two factors with three levels in the first factor and two levels in the second factor. Note that each of the six cells separated by dot-dash lines contains exactly one point.  By contrast, in the six-point randomly generated Latin hypercube design displayed in Figure~\ref{fig:lhsb}, two out of six such cells do not contain any point.

Orthogonal arrays have been used directly to carry out computer experiments; see, for example,  \cite{JosephHungSud}.  Compared with orthogonal arrays, OA-based Latin hypercubes are more  favorable for computer experiments.  When projected onto lower dimensions, the design points in orthogonal arrays often have replicates. This is
undesirable  at the early stages of  experimentation  when relatively few factors  are likely to be important.

The construction of OA-based Latin  hypercubes depends on the existence of orthogonal arrays.
For the existence results of orthogonal arrays, see, for example, \cite{hedayat1999orthogonal} and \cite{mukerjee2006modern}. A library of orthogonal arrays is freely available on the N.J.A.\ Sloane website and the MktEx macro using the software SAS \cite[]{kuhfeld2009orthogonal}.  It should be noted that, for certain given run sizes and numbers of factors,
orthogonal arrays of different numbers of levels  and different strengths may be available. For instance,   an $\OA(16,4^5,2)$, an $\OA(16,2^5,4)$ and an $\OA(16,2^5,2)$ all produce OA-based Latin hypercubes of 16 runs for 5 factors. However, orthogonal arrays with more levels and/or higher strength are  preferred because they lead to designs with better projection space-filling properties.

Given an orthogonal array, the construction of \cite{tang1993orthogonal} can produce many OA-based Latin hypercubes. There arises the problem of choosing a preferable OA-based Latin hypercube.
\cite{leary2003optimal} presented an algorithm for finding optimal OA-based Latin hypercubes  that minimize (\ref{eq:aefunc})
using the Euclidean distance between design points.
The optimization was performed via the simulated annealing algorithm \cite[]{morris1995exploratory} and  the columnwise-pairwise algorithm \cite[]{li1997columnwise}.

Recall the problem of estimating the mean $\mu$ in (\ref{eq:mu}) of a known function $y=f(\bx)$ using a design with $n$ points $\bx_1,\ldots, \bx_n$ in Section~\ref{subsec:intro}.  Latin hypercube sampling stratifies all univariate margins simultaneously and thus  achieves a variance reduction compared with simple random sampling, as quantified in  Theorem~\ref{theo:var2}.  Theorem~\ref{theo:oalhs} below from \cite{tang1993orthogonal} shows that a further variance reduction is achieved by OA-based Latin hypercube sampling.

\begin{theorem}\label{theo:oalhs}
Suppose that $f$ is continuous on $[0,1]^k$. Let $\hat{\mu}_{oalhs}$ denote the $\hat{\mu}$ in (\ref{eq:muhat}) with a randomly selected OA-based Latin hypercube design of $n$ points. Then we have that
$$\hbox{Var}(\hat{\mu}_{oalhs}) = \frac{1}{n} \hbox{Var}[f(\bx)] - \frac{1}{n} \sum_{j=1}^k \hbox{Var}[f_j(x_j)] - \frac{1}{n} \sum_{i<j}^k\hbox{Var}[f_{ij}(x_i,x_j)] + o(\frac{1}{n}), $$
\noindent where $x_j$ is the $j$th input of $\bx$, $f_j(x_j) = \hbox{E}[f(\bx)|x_j]-\mu$, and $f_{ij}(x_i,x_j) =
\hbox{E}[f(\bx)|x_i,x_j]- \mu -f_i(x_i)-f_j(x_j)$.
\end{theorem}

To better understand  this result, we  write
$$f(\bx) = \mu + \sum_{j=1}^k f_j(x_j) +
\sum_{i<j}^kf_{ij}(x_i,x_j) + r(\bx),$$
\noindent where the terms on the right side of the equation are
uncorrelated with each other. Thus, the variance decomposition of the function $f$ is
$$ \hbox{Var}[f(\bx)] = \sum_{j=1}^k \hbox{Var}[f_j(x_j)] + \sum_{i<j}^k \hbox{Var} [f_{ij}(x_i,x_j)] + \hbox{Var}[r(\bx)]. $$
We see that Latin hypercube sampling achieves a variance reduction by removing the variances of the main effects $f_j(x_j)$ from $\hbox{Var}[f(\bx)]$, and OA-based Latin hypercube sampling further removes the variances of the interactions $f_{ij}(x_i,x_j)$.

We conclude this section by mentioning that randomized orthogonal arrays proposed by \cite{owen1992central} also enjoy good space-filling properties in the low-dimensional projections.  Results similar to Theorem~\ref{theo:oalhs} are given in \cite{owen1992central}.

\subsection{Orthogonal and nearly orthogonal Latin hypercube designs}\label{subsec:orth}

This section discusses the properties and constructions of  Latin hypercube designs that have zero or small column-wise correlations in all two-dimensional projections. Such designs are called orthogonal and nearly orthogonal Latin hypercube designs.
Orthogonal Latin hypercube designs are directly useful in fitting data using main effects models because they allow uncorrelated estimates of linear main effects. Another rationale of obtaining orthogonal or nearly orthogonal Latin hypercube designs is that  they  may not be space-filling, but space-filling designs should be orthogonal or nearly orthogonal.
Thus we can search for space-filling designs within the class of orthogonal or nearly orthogonal Latin hypercube designs. Other justifications are given in \cite{iman1982distribution}, \cite{owen1994controlling}, \cite{tang1998selecting}, \cite{joseph2008orthogonal}, among others.

Extensive research has been carried out on the construction of orthogonal or nearly orthogonal Latin hypercube designs. \cite{ye1998orthogonal} initiated this line of research and constructed orthogonal Latin hypercubes with  $n=2^m$ or $2^m+1$ runs and  $k=2m-2$ factors where $m \geq 2$.
Ye's construction was extended by \cite{cioppa2007efficient} to obtain more columns for  given run sizes.
\cite{steinberg2006construction} constructed orthogonal Latin hypercubes of the run sizes $n=2^{2^m}$ by rotating groups of factors in two-level $2^{2^m}$-run
regular fractional factorial designs. This idea was generalized by
\cite{pang2009construction} who constructed orthogonal Latin hypercubes of $p^{2^m}$ runs and up to $(p^{2^m}-1)/(p-1)$ factors by rotating groups of factors in $p$-level $p^{2^m}$-run
regular  factorial designs, where $p$ is a prime. \cite{lin2008} obtained orthogonal Latin hypercube designs of small run sizes ($n \leq 20$) through an algorithm that adds columns sequentially to an existing design.
More recently, various methods have been proposed
to construct orthogonal Latin hypercubes of more flexible run sizes and with large factor-to-run-size ratios.
Here we review constructions in \cite{lin2009construction}, \cite{sun2009construction}, and \cite{lin2010new}. These methods are general, simple to implement, and cover the results in Table~\ref{table:large}.
For other methods, see \cite{georgiou2009orthogonal}, \cite{sun2010construction}, and \cite{yangconstruction}.

We first give some   notation and background.
A vector $\ba=(a_1,\ldots,a_n)$ is said to be balanced if its distinct values have equal frequency.
For an $n_1 \times k_1$ matrix $\bA$ and an $n_2 \times k_2$ matrix $\bB$, their Kronecker product $\bA \otimes \bB$   is the $(n_1n_2)
\times (k_1k_2)$  matrix
\begin{center}
$\bA \otimes \bB =\left [
\begin{array}{cccc}
  a_{11}\bB   &  a_{12}\bB   &   \ldots  &   a_{1k_1}\bB  \\
  a_{21}\bB   &  a_{22}\bB   &   \ldots  &   a_{2k_1}\bB  \\
  \vdots    &  \vdots    &   \ddots  &   \vdots   \\
  a_{n_11}\bB &  a_{n_12}\bB &   \ldots  &   a_{n_1k_1}\bB
\end{array}\right]$
\end{center}
\noindent with $a_{ij}\bB$ itself being an $n_2 \times k_2$ matrix. For an $n \times k$ design or matrix $\bD=(d_{ij})$, define its correlation matrix to be a $k \times k$ matrix $\bR(\bD) = (r_{ij})$ with
\begin{equation}\label{eq:rho}
r_{ij}=\frac{\sum_{m=1}^n(d_{mi}-\bar{d}_i)(d_{mj}-\bar{d}_j)}
{\sqrt{\sum_m(d_{mi}-\bar{d}_i)^2
\sum_m(d_{mj}-\bar{d}_j)^2}}
\end{equation}
\noindent representing the correlation between the $i$th and $j$th columns of $\bD$, where $\bar{d}_i=n^{-1}\sum_{m=1}^n d_{mi}$ and $\bar{d}_j=n^{-1}\sum_{m=1}^n d_{mj}$. A design or matrix $\bD$ is  column-orthogonal if $\bR(\bD)$ is an identity matrix.
A design or matrix $\bD=(d_{ij}) $ is orthogonal if  it is balanced and column-orthogonal.

To assess near orthogonality of design $\bD$, \cite{bingham2009orthogonal} introduced two measures,  the maximum correlation
$\rho_M(\bD)=\hbox{max}_{i < j}|r_{ij}|$ and the average
squared correlation  $\rho^2_{ave}(\bD)=\sum_{i<j}r^2_{ij}/[(k(k-1)/2]$, where $r_{ij}$ is defined as in (\ref{eq:rho}).
Smaller values of $\rho_M(\bD)$ and $\rho^2_{ave}(\bD)$ imply near
orthogonality. Obviously, if $\rho_M(\bD)=0$ or $\rho^2_{ave}(\bD)=0$, then an orthogonal design is obtained. For a concise presentation, we use $\OLH(n,k)$
to denote an orthogonal Latin hypercube of $n$ runs for $k$ factors. \cite{lin2010new} established the following theorem on the existence  of orthogonal Latin hypercubes.
\begin{theorem}\label{theo:exist}
There exists an orthogonal Latin hypercube of $n \geq 4$ runs with
more than one factor if and only if $n \neq 4m + 2$ where
$m$ is an integer.
\end{theorem}

Theorem~\ref{theo:exist} says that a Latin hypercube of run size 2, 3, 6, 10, 14, $\ldots$ cannot be orthogonal. For smaller run sizes, this can be readily verified by exhaustive computer  search.
When orthogonal Latin hypercubes of certain run sizes exist, we want to construct such designs with as many columns as possible. We review three general construction methods. To generate design points in the region $[0,1]^k$ from a Latin hypercube, one can use (\ref{eq:dij}) with $u_{ij}=0.5$, which corresponds to the midpoints of the cells.


\subsubsection{A construction method based on an orthogonal array and a small orthogonal Latin hypercube}\label{subsubsec:orth1}
\cite{lin2009construction} constructed a large orthogonal, or nearly orthogonal, Latin hypercube by coupling an orthogonal array of index unity with a small orthogonal, or nearly orthogonal, Latin hypercube. Let $\bB $  be an $n \times q$ Latin hypercube, where as in Section~\ref{subsec:intro}, the levels are $-(n-1)/2, -(n-3)/2, \ldots, (n-3)/2, (n-1)/2$. Then the elements in every column of $\bB$ add up to zero
while the sum of squares of these elements is $n(n^2-1)/12$. Thus the correlation matrix whose elements are defined as in (\ref{eq:rho}) is
\begin{equation}\label{eq:rb}
\bR(\bB) = \big [ \frac{1}{12}n(n^2-1)\big ]^{-1}\bB^\prime \bB.
\end{equation}
Let $\bA$ be an orthogonal array $\OA(n^2, n^{2f}, 2)$. The construction proposed by \cite{lin2009construction} proceeds as follows.
\begin{myindentpar}{0.5cm}
{\em Step I}: Let $b_{ij}$ be the $(i,j)$th element of $\bB$ introduced above. For $1 \leq j \leq q$, obtain an $n^2 \times (2f)$
matrix $\bA_j$ from $\bA$ by replacing the symbols $1, 2, \ldots, n$ in
the latter by $b_{1j}, b_{2j}, \ldots, b_{nj}$ respectively, and
then partition $\bA_j$ as $\bA_j$ = [$\bA_{j1}, \ldots, \bA_{jf}$], where
each of $\bA_{j1}, \ldots, \bA_{jf}$ has two columns.

{\em Step II}: For $1 \leq j \leq q$, obtain the $n^2 \times (2f)$
matrix $\bL_j$ = [$\bA_{j1}\bV, \ldots, \bA_{jf}\bV$] , where
\[
{
\renewcommand{\arraystretch}{0.6}
\bV = \left[
\begin{array}{rr}
1 & -n \\
n & 1 \\
\end{array}
\right]. }
\]

{\em Step III}: Obtain the matrix $\bL$ = [$\bL_{1}, \ldots,
\bL_{q}$], of order $N \times k$, where $N=n^2$ and $k=2qf$.
\end{myindentpar}

For $q=1$,   the above construction is
equivalent to that in  \cite{pang2009construction}. However, by Theorem~\ref{theo:exist}, we have $q \geq 2$ when $n$ is not equal to 3 or $4m+2$ for any non-negative integer $m$. Thus,
the above method provides orthogonal or nearly orthogonal Latin hypercubes with an appreciably larger number of factors as
compared to the method in  \cite{pang2009construction}. For example, \cite{pang2009construction} obtained $\OLH(25,6)$, $\OLH(49,8)$, $\OLH(81,40)$, $\OLH(121,12)$ and $\OLH(169,14)$ while the above construction produces
 $\OLH(25,12)$, $\OLH(49,24)$, $\OLH(81,50)$, $\OLH(121,84)$ and $\OLH(169,84)$.

Theorem~\ref{theo:lmt} below shows how the correlation matrix of the large Latin hypercube $\bL$ depends on that of the small Latin hypercube $\bB$.

\begin{theorem}\label{theo:lmt}
For the matrix $\bL$ constructed from $\bB$ in the above steps I, II and III, we have\\
(i) the matrix $\bL$ is a Latin hypercube, and\\
(ii) the correlation matrix of $\bL$ is given by
$$ \bR(\bL) = \bR(\bB) \otimes \bI_{2f},$$
\noindent where $\bR(\bB)$, given in (\ref{eq:rb}), is the correlation matrix of a Latin hypercube $\bB$, $\bI_{2f}$ is the identity matrix of order $2f$ and
$\otimes$ denotes Kronecker product.
\end{theorem}

\begin{corollary}\label{cor:lmt}
If $\bB$ is an orthogonal Latin hypercube, then so is $\bL$. In general, the maximum correlation and average squared correlation of $\bL$ are given by
$$ \rho_M(\bL) = \rho_M(\bB) \ \  \hbox{ and } \ \  \rho_{ave}^2(\bL) = \frac{q-1}{2qf-1}
\rho_{ave}^2(\bB).$$
\end{corollary}

Corollary~\ref{cor:lmt} reveals that the large Latin hypercube $\bL$
inherits the exact or near orthogonality of the small Latin hypercube $\bB$. As a result, the effort for searching for large orthogonal or nearly
orthogonal Latin hypercubes can be focused on finding small orthogonal or nearly orthogonal Latin hypercubes which are easier to obtain via some general  efficient robust optimization algorithms such as simulated annealing and genetic algorithms, by minimizing $\rho_{ave}^2$
or $\rho_M$.

Example~\ref{exam:oa} below illustrates the actual construction of some orthogonal Latin hypercubes using the method of  \cite{lin2009construction}. Example~\ref{exam:noa} is devoted to the construction of a nearly orthogonal Latin hypercube.

\begin{example}\label{exam:oa}
Let $n$ be a prime or prime power for which an $\OA(n^2,n^{n+1},2)$ exists \cite[]{hedayat1999orthogonal}. For instance, consider $n=5,7,8,9,11$.
Now if we take $\bB$ to be an $\OLH(5,2)$, an $ \OLH(7,3)$, an $\OLH(8,4)$, an $\OLH(9,5)$, or an $\OLH(11,7)$, as displayed in Table~\ref{table:olhsmall} and take $\bA$ respectively to be
an $\OA(25,5^6,2)$, an $\OA(49,7^8,2)$, an $\OA(64,8^8,2)$, an $\OA(81,9^{10},2)$, or an $\OA(121,11^{12},2)$, then the construction described in this section provides an $\OLH(25,12)$, an $\OLH(49,24)$, an $\OLH(64,32)$, an $\OLH(81,50)$, or an $\OLH(121,84)$, respectively.
\begin{table}[!htb]
\begin{center}
\caption{Orthogonal Latin hypercubes $\OLH$($5,2$), $\OLH$($7,3$), $\OLH$($8,4$), $\OLH$($9,5$) and $\OLH$($11,7$)}
\scalebox{0.9}{
\begin{tabular}{rrrrrrrrrrrrrrrrrr}
 \multicolumn{3}{c}{$\OLH(5,2)$} && &\multicolumn{3}{c}{$\OLH(7,3)$}&&& &\multicolumn{4}{c}{$\OLH(8,4)$}  \\
   1&   -2 & $\quad$ &  $\quad$ & $\quad$ & -3&    3&    2 & &  \ &  \ &  0.5 & -1.5  &  3.5  & 2.5   \\
   2&    1 & \ &  \ & \ & -2&    0&   -3 & \ &  \ &  \ &  1.5 &  0.5  &  2.5  &-3.5   \\
   0&    0 & \ &  \ & \ & -1&   -2&   -1 & \ &  \ &  \ &  2.5 & -3.5  & -1.5  &-0.5   \\
  -1&    2 & \ &  \ & \ &  0&   -3&    1 & \ &  \ &  \ &  3.5 &  2.5  & -0.5  & 1.5   \\
  -2&   -1 & \ &  \ & \ &  1&   -1&    3 & \ &  \ &  \ & -3.5 & -2.5  &  0.5  &-1.5   \\
    &      & \ &  \ & \ &  2&    1&   -2 & \ &  \ &  \ & -2.5 &  3.5  &  1.5  & 0.5   \\
    &      & \ &  \ & \ &  3&    2&    0 & \ &  \ &  \ & -1.5 & -0.5  & -2.5  & 3.5   \\
    &      & \ &  \ & \ &   &     &      & \ &  \ &  \ & -0.5 &  1.5  & -3.5  &-2.5   \\
    &&&&&&&&&&&&&&\\
  \multicolumn{5}{c}{$\OLH(9,5)$} & &&&
  \multicolumn{7}{c}{$\OLH(11,7)$}\\
 -4&-2&    0&   -3&    3& \  &  \ &  \ &  -5 &  -4&   -5 &  -5&   -3 &   0&   0\\
 -3& 4&    2&    1&   -2& \  &  \ &  \ &  -4 &   2&   -1 &   3&    4 &   5&   4\\
 -2&-3&   -4&   -1&   -3& \  &  \ &  \ &  -3 &  -2&    4 &   5&   -4 &  -2&  -1\\
 -1& 3&   -2&    3&    4& \  &  \ &  \ &  -2 &   3&   -3 &   4&    1 &  -4&  -2\\
  0&-4&    4&    4&    0& \  &  \ &  \ &  -1 &   4&    2 &  -4&    3 &   2&  -4\\
  1& 2&   -1&    0&   -4& \  &  \ &  \ &   0 &  -5&    5 &  -2&    5 &  -3&   2\\
  2& 0&    3&   -2&   -1& \  &  \ &  \ &   1 &   5&    3 &  -3&   -5 &  -1&   5\\
  3& 1&    1&   -4&    2& \  &  \ &  \ &   2 &  -1&    1 &   1&   -2 &   3&  -5\\
  4&-1&   -3&    2&    1& & & &   3 &   0&    0 &  -1&    0 &   1&  -3\\
   &  &     &     &     & & & &   4 &   1&   -4 &   0&    2 &  -5&   1\\
   &  &     &     &     & & & &   5 &  -3&   -2 &   2&   -1 &   4&   3\\
\end{tabular}}\label{table:olhsmall}
\end{center}
\end{table}

\end{example}

\begin{example}\label{exam:noa}
Through a computer search, \cite{lin2009construction} found a nearly orthogonal Latin hypercube with 13 rows and 12 columns, as given in
Table~\ref{Table:nolh}. This Latin hypercube has $\rho_{ave}= 0.0222$ and $\rho_M=0.0495$.
Together with an $\OA(13^2,13^{14},2)$, the above procedure provides a nearly orthogonal Latin hypercube  of 169 runs and 168 factors with $\rho_{ave} = 0.0057$  and $\rho_M=0.0495$, by Corollary~\ref{cor:lmt}.
\begin{table}[!t]
\begin{center}
\caption{A nearly orthogonal Latin hypercube with 13 rows and 12 columns }
\scalebox{0.86}{
\begin{tabular}{c}
$\begin{array}{rrrrrrrrrrrr}
 -6  &  -6  &  -5 &   -4  &  -5  &  -2 &    2 &    1 &   -3 &   -2 &   -1 &   -2  \\
 -5  &   5  &   3 &   -5  &   3  &   4 &   -6 &    0 &   -4 &    1 &   -3 &   -1  \\
 -4  &   2  &  -4 &    1  &   2  &   6 &    5 &   -5 &    6 &    0 &    1 &    1  \\
 -3  &   1  &   2 &    4  &  -6  &   1 &   -2 &    6 &    2 &    3 &    2 &    6  \\
 -2  &  -2  &   6 &   -3  &   6  &  -5 &    3 &    4 &    4 &   -3 &    3 &    0  \\
 -1  &  -5  &   4 &    6  &   1  &  -1 &    0 &   -4 &    0 &    6 &   -5 &   -3  \\
  0  &   6  &   0 &    3  &  -4  &  -6 &   -3 &   -3 &    3 &   -5 &    0 &   -4  \\
  1  &   0  &  -3 &    5  &   5  &   0 &    1 &    2 &   -5 &   -6 &   -4 &    5  \\
  2  &  -1  &  -6 &    0  &   4  &  -4 &   -5 &   -2 &   -1 &    5 &    6 &    2  \\
  3  &   4  &   1 &    2  &  -1  &   2 &    6 &    3 &   -6 &    2 &    5 &   -6  \\
  4  &  -4  &   5 &   -2  &  -3  &   3 &   -1 &   -6 &   -2 &   -4 &    4 &    3  \\
  5  &   3  &  -1 &   -6  &  -2  &  -3 &    4 &   -1 &    1 &    4 &   -6 &    4  \\
  6  &  -3  &  -2 &   -1  &   0  &   5 &   -4 &    5 &    5 &   -1 &   -2 &   -5
\end{array}$
\end{tabular}}\label{Table:nolh}
\end{center}
\end{table}
\end{example}

Before ending this section, we comment on the projection space-filling property of Latin hypercubes built above using  a Latin hypercube $\bB$ and an orthogonal array $\bA$. Any pair of columns  obtained using different columns of   $\bA$ retains the two-dimensional projection property of   $\bA$. When projected to those pairs of columns associated with the same column of $\bA$, the design points form clusters and these clusters are spread out as the corresponding two columns of $\bB$.

\subsubsection{A recursive construction method}\label{subsubsec:orth2}
Orthogonal Latin hypercubes allow uncorrelated estimates of main effects in a main-effect regression model. \cite{sun2009construction} extended the concept of orthogonal Latin hypercubes for second-order polynomial models.

For a design $\bD$ with columns $\bd_1,\ldots,\bd_k$, let $\tilde{\bD}$ be the $n\times [k(k+1)/2]$
matrix whose columns consist of all possible products
$\bd_i\odot \bd_j$, where $\odot$ denotes the element-wise product of
vectors $\bd_i$ and $\bd_j$, $i=1,\ldots,k$, $j=1,\ldots,k$ and $i \leq j$. Define the correlation matrix between $\bD$ and $\tilde{\bD}$ to
be
\begin{eqnarray}\label{eq:rhomat2}
\bR(\bD, \tilde{\bD})=\left(
\begin{array}{rrrr}
r_{11} & r_{12} & \ldots & r_{1q}\\
r_{21} & r_{22} & \ldots & r_{2q}\\
\vdots & \vdots  &\ddots &\vdots\\
r_{k1} & r_{k2} & \ldots & r_{kq}\\
\end{array}
\right),
\end{eqnarray}
where $q=k(k+1)/2$ and $r_{ij}$ is the correlation between the $i$th column of
$\bD$ and the $j$th column of $\tilde{\bD}$.  A second-order orthogonal Latin hypercube $\bD$ has the properties: (a) the correlation matrix $\bR(\bD)$
is an identity matrix, and (b) $\bR(\bD,\tilde{\bD})$ in (\ref{eq:rhomat2})
is a zero matrix.

\cite{sun2009construction} proposed the following procedure for constructing second-order orthogonal Latin hypercubes of $2^{c+1}+1$ runs in $2^c$ factors for any integer $c \geq 1$.
Throughout this section, let $\bX^*$ represent the matrix obtained by switching the signs in the top half of the matrix $\bX$ with an even number of rows.
\begin{myindentpar}{0.5cm}
{\em Step I}: For $c=1$, let
\begin{equation*}
\bS_1=\left(\begin{array}{rr}
1&1\\
1&-1 \end{array}\right) \ \mbox{\ and\ } \ \  \bT_1=\left( \begin{array}{rr}
1&2\\
2&-1\end{array}\right).
\end{equation*}

{\em Step II}: For an integer $c \geq 2$, define
\begin{equation*} \label{T_c}
\bS_c=\left(\begin{array}{cc}
\bS_{c-1}&-\bS^*_{c-1}\\
\bS_{c-1}& \bS^*_{c-1} \end{array}\right)\  \mbox{\ and\ }\
\bT_c=\left( \begin{array}{cc}
\bT_{c-1}&-(\bT^*_{c-1}+2^{c-1}\bS^*_{c-1})\\
\bT_{c-1}+2^{c-1}\bS_{c-1}&\bT^*_{c-1}\end{array}\right).
\end{equation*}

{\em Step III}: Obtain a $(2^{c+1}+1)\times 2^c$ Latin hypercube $\bL_c$ as
\begin{equation*}\label{eq:L_c}
\bL_c = \left(
\begin{array}{r}
\bT_c\\
\bzero_{2^c}\\
-\bT_c\\
\end{array}
\right),
\end{equation*}
\noindent where $\bzero_{2^c}$ denotes  a zero row vector of length $2^c$.
\end{myindentpar}

\begin{example}
A second-order orthogonal Latin hypercube of 17 runs for 8 factors constructed using the above procedure  is given by
{\small
$${ \left(\begin{array}{rrrrrrrr}
   1  &  2  &  3  &  4  &  5 &   6  &  7&    8    \\ [-6pt]
   2  & -1  & -4  &  3  &  6 &  -5  & -8&    7    \\ [-6pt]
   3  &  4  & -1  & -2  & -7 &  -8  &  5&    6    \\ [-6pt]
   4  & -3  &  2  & -1  & -8 &   7  & -6&    5    \\ [-6pt]
   5  &  6  &  7  &  8  & -1 &  -2  & -3&   -4    \\ [-6pt]
   6  & -5  & -8  &  7  & -2 &   1  &  4&   -3    \\ [-6pt]
   7  &  8  & -5  & -6  &  3 &   4  & -1&   -2    \\ [-6pt]
   8  & -7  &  6  & -5  &  4 &  -3  &  2&   -1    \\ [-6pt]
   0  &  0  &  0  &  0  &  0 &   0  &  0&    0    \\ [-6pt]
  -1  & -2  & -3  & -4  & -5 &  -6  & -7&   -8    \\ [-6pt]
  -2  &  1  &  4  & -3  & -6 &   5  &  8&   -7    \\ [-6pt]
  -3  & -4  &  1  &  2  &  7 &   8  & -5&   -6    \\ [-6pt]
  -4  &  3  & -2  &  1  &  8 &  -7  &  6&   -5    \\ [-6pt]
  -5  & -6  & -7  & -8  &  1 &   2  &  3&    4    \\ [-6pt]
  -6  &  5  &  8  & -7  &  2 &  -1  & -4&    3    \\ [-6pt]
  -7  & -8  &  5  &  6  & -3 &  -4  &  1&    2    \\ [-6pt]
  -8  &  7  & -6  &  5  & -4 &   3  & -2&    1    \\
   \end{array}\right).}$$ }
\end{example}

\cite{sun2010construction} further constructed second-order orthogonal Latin hypercubes of $2^{c+1}$ runs in $2^c$ factors by modifying Step III in the above procedure. In Step III, let $\bH_c = \bT_c - \bS_c/2$ and obtain $\bL_c$ as
\begin{equation*}\label{eq:L_c2}
\bL_c = \left(
\begin{array}{r}
\bH_c \\
-\bH_c\\
\end{array}
\right).
\end{equation*}
\noindent Then $\bL_c$ is a second-order orthogonal Latin hypercube of $2^{c+1}$ runs in $2^c$ factors.

\subsubsection{A construction method based on small orthogonal designs and small orthogonal Latin hypercubes}\label{subsubsec:orth3}
This section reviews the construction from \cite{lin2010new} for constructing orthogonal and nearly orthogonal Latin hypercubes. All the proofs can be found in \cite{lin2010new}. Let $\bA=(a_{ij})$
  be an $n_1 \times k_1$ matrix with entries $a_{ij}=\pm 1$,
$\bB=(b_{ij})$ be an ${n_2 \times k_2}$ Latin hypercube,
$\bE=(e_{ij})$
  be an ${n_1 \times k_1}$ Latin hypercube,
and $\bF=(f_{ij})$ be an ${n_2 \times k_2}$ matrix with entries $d_{ij}=\pm 1$.
\cite{lin2010new}  construct designs via
\begin{equation}\label{eq:kron}
\bL=\bA\otimes \bB + n_2 \bE\otimes \bF.
\end{equation}
The resulting design $\bL$ in (\ref{eq:kron}) has $n=n_1n_2$ runs and
$k=k_1k_2$ factors, and becomes an orthogonal Latin hypercube, if certain conditions on $\bA, \bB, \bE, \bF$ are met.

\begin{theorem}\label{theo:kron}
Design $\bL$ in (\ref{eq:kron}) is an orthogonal Latin hypercube if \\
(i) $\bA$ and $\bF$ are column-orthogonal matrices of $\pm 1$, \\
(ii) $\bB$ and $\bE$ are orthogonal Latin hypercubes,  \\
(iii) at least one of the two, $\bA^\prime \bE=\bzero$ and $\bB^\prime \bF=\bzero$, is true, and \\
(iv)
at least one of the following two conditions is true: \\
\indent
(a) $\bA$ and $\bE$ satisfy that for any $i$, if $p_1$ and $p_2$ are such that
$e_{p_1i} = - e_{p_2i}$, then $a_{p_1i} = a_{p_2i}$; \\
\indent (b)
$\bB$ and $\bF$ satisfy that
for any $j$, if $q_1$ and $q_2$ are such that
$b_{q_1j} = - b_{q_2j}$, then $f_{q_1j} = f_{q_2j}$.
\end{theorem}

Condition  (iv) in Theorem~\ref{theo:kron} is needed for $\bL$ to be a Latin hypercube.
To make $\bL$ orthogonal, conditions (i), (ii) and (iii)
are necessary. Choices for $\bA$ and $\bF$ include Hadamard matrices and orthogonal arrays with levels $\pm 1$ (see Chapter~9).  Because of the orthogonality of $\bA$ and $\bF$, $n_1$ and $n_2$ must be equal to two or multiples of four \cite[][p.\ 33]{DeyMukerjee}.
Theorem~\ref{theo:kron} requires designs $\bB$ and $\bE$ to be orthogonal Latin hypercubes. All
known orthogonal Latin hypercubes of run sizes that are two or multiples of four can be used.
As a result, Theorem~\ref{theo:kron} can be used to construct a vast number of orthogonal Latin hypercubes of $n=8k$ runs.  Example~\ref{ex:kron} illustrates the use of Theorem~\ref{theo:kron}.

\begin{example}\label{ex:kron}
Consider constructing an orthogonal Latin hypercube of 32 runs. Let $\bA=(1,1)^\prime$, $\bB$ be the $16 \times 12$ orthogonal Latin hypercube in Table~\ref{table:lhd16}, $\bE=(1/2,-1/2)^\prime$, and $\bF$ be a matrix obtained by taking any 12 columns from a Hadamard matrix of order 16.
By Theorem~\ref{theo:kron},  $\bL$ in (\ref{eq:kron}) with the chosen $\bA,\bB,\bE,\bF$ constitutes a $32 \times 12$ orthogonal Latin hypercube.
\begin{table}[!htb]
\caption{A $16 \times 12$ orthogonal Latin hypercube from \cite{steinberg2006construction}}
\label{table:lhd16}
\begin{center}
\scalebox{0.9}{
\begin{tabular}{c}
$\bB=\frac{1}{2}
\left(
\begin{array}{rrrrrrrrrrrrrrrr}
-15&  5&  9 &  -3&   7&  11& -11&   7 &  -9&   3& -15& 5   \\
-13&  1&  1 &  13&  -7& -11&  11&  -7 &  -1& -13& -13& 1   \\
-11&  7& -7 & -11&  13&  -1&  -1& -13 &   9&  -3&  15& -5  \\
-9 & 3 & -15&   5& -13&   1&   1&  13 &   1&  13&  13& -1  \\
-7 &-11&  11&  -7&  11&  -7&   7&  11 &   5&  15&  -3& -9   \\
-5 &-15&   3&   9& -11&   7&  -7& -11 &  13&  -1&  -1& -13  \\
-3 &-9 &  -5& -15&   1&  13&  13&  -1 &  -5& -15&   3& 9    \\
-1 &-13& -13&   1&  -1& -13& -13&   1 & -13&   1&   1& 13   \\
1  & 13&  13&  -1&  -9&   3& -15&   5 &  11&  -7&   7& 11  \\
3  & 9 &   5&  15&   9&  -3&  15&  -5 &   3&   9&   5& 15  \\
5  & 15&  -3&  -9&  -3&  -9&  -5& -15 & -11&   7&  -7& -11 \\
7  & 11& -11&   7&   3&   9&   5&  15 &  -3&  -9&  -5& -15 \\
9  & -3&  15&  -5&  -5& -15&   3&   9 &  -7& -11&  11& -7  \\
11 & -7&   7&  11&   5&  15&  -3&  -9 & -15&   5&   9& -3  \\
13 & -1&  -1& -13& -15&   5&   9&  -3 &   7&  11& -11& 7   \\
15 & -5&  -9&   3&  15&  -5&  -9&   3 &  15&  -5&  -9& 3   \\
\end{array}\right)$
\end{tabular}}
\end{center}
\end{table}
\end{example}

When $n_1 = n_2$, a stronger result than Theorem~\ref{theo:kron}, as given in Theorem~\ref{prop:kron},  can be established.
It provides orthogonal Latin hypercubes with more columns than those in Theorem~\ref{theo:kron}.

\begin{theorem}\label{prop:kron}
If $n_1=n_2$ and $\bA$, $\bB$, $\bE$, and $\bF$
are chosen according to Theorem~\ref{theo:kron},
then design ($\bL, \bU$) is an orthogonal Latin hypercube with $2k_1k_2$
factors, where $\bL$ is as in Theorem~\ref{theo:kron} and
$\bU=-n_1\bA \otimes \bB + \bE \otimes \bF$.
\end{theorem}

\begin{example}
To construct  orthogonal Latin hypercubes of 64 runs, let $n_1=n_2=8$ and take
$${ \small \bA=\bF=\left(\begin{array}{rrrr}
1&  1&  1&  1\\[-6pt]
1&  1& -1& -1\\[-6pt]
1& -1&  1& -1\\[-6pt]
1& -1& -1&  1\\[-6pt]
1&  1&  1&  1\\[-6pt]
1&  1& -1& -1\\[-6pt]
1& -1&  1& -1\\[-6pt]
1& -1& -1&  1\\ \end{array}
\right)}, \hbox{ and } \ {  \small
\bB=\bE= \frac{1}{2} \left(\begin{array}{rrrr}
1 & -3 &7 &5 \\       [-6pt]
3 &1& 5 &-7  \\       [-6pt]
5 &-7& -3& -1\\        [-6pt]
7 &5& -1& 3  \\        [-6pt]
-1& 3& -7 &-5\\       [-6pt]
-3 &-1& -5& 7\\       [-6pt]
-5 &7 &3& 1\\         [-6pt]
-7& -5& 1& -3\\       \end{array}
\right)}.$$
\noindent Then design $(\bL,\bU)$ in Theorem~\ref{prop:kron} is a $64 \times 32$ orthogonal Latin hypercube.
\end{example}

\begin{theorem}\label{theo:bound}
Suppose that an $\OLH(n,k)$ is available,
where $n$ is a multiple of 4 such that a Hadamard matrix
of order $n$ exists. Then  the following orthogonal Latin hypercubes,
an $\OLH$($2n,k$),
an $\OLH$($4n,2k$),
an $\OLH$($8n,4k$) and
an $\OLH$($16n,8k$),
can all be constructed.
\end{theorem}

We give a sketch of the proof for Theorem~\ref{theo:bound}. The proof provides the actual construction of these orthogonal Latin hypercubes.
The theorem results from an application of Theorem~\ref{theo:kron} and the use of orthogonal designs in Table~\ref{table:od}.
Note that each of the four matrices  in Table~\ref{table:od} can be written as $(\bX^\prime,-\bX^\prime)^\prime$.
From such an $\bX$, define $\bS$ to be the matrix obtained by choosing $x_i=1$ for all $i$'s. Let $\bA = (\bS^\prime,\bS^\prime)^\prime$. Further let $\bE$ be an orthogonal Latin hypercube derived from a matrix in Table~\ref{table:od} by letting $x_i=(2i-1)/2$ for $i=1,\ldots,n/2$. Now we choose $\bB$ to be a given $\OLH(n,k)$ and $\bF$ be the matrix obtained by taking any $k$ columns from a Hadamard matrix order $n$. Such matrices $\bA$, $\bB$, $\bE$, and $\bF$ meet conditions (i), (ii), (iii) and (iv) in Theorem~\ref{theo:kron}, from which  Theorem~\ref{theo:bound} follows.

\begin{table}[t]
\caption{Four orthogonal designs}\label{table:od}
\begin{center}
\scalebox{0.78}{
\begin{tabular}{cccccccccccc}
\hline \hline
\multicolumn{12}{c}{$n$}\\
\hline
2 &  \multicolumn{2}{c}{$4$} &\multicolumn{3}{c}{$8$}&  \multicolumn{6}{c}{$16$} \\
\hline
 $\begin{array}{r}
  x_1 \\
 -x_1  \\
   \\
 \\
 \\
 \\
 \\
 \\
 \\
 \\
 \\
 \\
 \\
 \\
 \\
 \\
\end{array}$ &  \multicolumn{2}{c}{
$\begin{array}{rr}
  x_1 & x_2\\
  x_2 & -x_1 \\
  -x_1 & -x_2 \\
  -x_2 & x_1\\
& \\                  & \\                  & \\
  & \\                  & \\                  & \\
  & \\                  & \\                  & \\
  & \\                  & \\
& \\
\end{array}
$}& \multicolumn{3}{c}{$\begin{array}{rrrr}
x_1 &-x_2& x_4 & x_3\\
x_2 & x_1 & x_3 & -x_4\\
x_3 & -x_4 & -x_2 & -x_1\\
x_4 & x_3 & -x_1 & x_2\\
-x_1 & x_2 & -x_4 & -x_3\\
-x_2 & -x_1 & -x_3 &x_4\\
-x_3 & x_4 & x_2 & x_1\\
-x_4 & -x_3 & x_1 & -x_2\\
    &     &      &    \\
    &     &      &    \\
    &     &      &    \\
    &     &      &    \\
    &     &      &    \\
    &     &      &    \\
    &     &      &    \\
    &     &      &    \\
\end{array}$}&  \multicolumn{6}{c}{
$\begin{array}{rrrrrrrr}
  x_1 & -x_2 & -x_4 & -x_3 & -x_8 & x_7&  x_5 &  x_6 \\
  x_2 &  x_1 & -x_3 &  x_4 & -x_7 &-x_8& -x_6 &  x_5 \\
  x_3 & -x_4 &  x_2 &  x_1 & -x_6 &-x_5&  x_7 & -x_8 \\
  x_4 &  x_3 &  x_1 & -x_2 & -x_5 & x_6& -x_8 & -x_7 \\
  x_5 & -x_6 & -x_8 &  x_7 &  x_4 & x_3& -x_1 & -x_2 \\
  x_6 &  x_5 & -x_7 & -x_8 &  x_3 &-x_4&  x_2 & -x_1 \\
  x_7 & -x_8 &  x_6 & -x_5 &  x_2 &-x_1& -x_3 &  x_4 \\
  x_8 &  x_7 &  x_5 &  x_6 &  x_1 & x_2&  x_4 &  x_3 \\
 -x_1 &  x_2 &  x_4 &  x_3 &  x_8 &-x_7& -x_5 & -x_6 \\
 -x_2 & -x_1 &  x_3 & -x_4 &  x_7 & x_8&  x_6 & -x_5 \\
 -x_3 &  x_4 & -x_2 & -x_1 &  x_6 & x_5& -x_7 &  x_8 \\
 -x_4 & -x_3 & -x_1 &  x_2 &  x_5 &-x_6&  x_8 &  x_7 \\
 -x_5 &  x_6 &  x_8 & -x_7 & -x_4 &-x_3&  x_1 &  x_2 \\
 -x_6 & -x_5 &  x_7 &  x_8 & -x_3 & x_4& -x_2 &  x_1 \\
 -x_7 &  x_8 & -x_6 &  x_5 & -x_2 & x_1&  x_3 & -x_4 \\
 -x_8 & -x_7 & -x_5 & -x_6 & -x_1 &-x_2& -x_4 & -x_3 \\
\end{array}$}\\
\hline
\end{tabular}}
\end{center}
\end{table}

Theorem~\ref{theo:bound} is a very powerful result.  By repeatedly applying Theorem~\ref{theo:bound}, one can obtain many infinite series of orthogonal
Latin hypercubes. For example,
starting with an $\OLH(12,6)$, we can obtain
an $\OLH(192,48)$, which can be used in turn to construct
an $\OLH(768,96)$ and so on. For another example,
an $\OLH(256,248)$ in \cite{steinberg2006construction}
can be used to construct
an $\OLH(1024,496)$, an $\OLH(4096,1984)$ and so on.

Another result from \cite{lin2010new} shows how the method in (\ref{eq:kron}) can be used to construct nearly orthogonal Latin hypercubes.

\begin{theorem}\label{prop:nearorth}
Suppose that   condition (iv) in Theorem~\ref{theo:kron} is satisfied so
that design $\bL$ in (\ref{eq:kron}) is a Latin hypercube.
If  conditions (i) and (iii) in Theorem~\ref{theo:kron} hold, we then have that\\
(i) $\rho_{ave}^2(\bL)=w_1\rho_{ave}^2(\bB)+w_2\rho_{ave}^2(\bE)$,
  and\\
(ii) $\rho_M(\bL)=\hbox{max}\{w_3\rho_M(\bB),
  w_4\rho_M(\bE)\}$, \\
\noindent where $w_1$, $w_2$, $w_3$ and $w_4$ are given by
  $w_1=(k_2-1)(n_2^2-1)^2/[(k_1k_2-1)(n^2-1)^2]$,
  $w_2=n_2^4(k_1-1)(n_1^2-1)^2/[(k_1k_2-1)(n^2-1)^2]$, $w_3=(n_2^2-1)/(n^2-1)$ and
$w_4=n_2^2(n_1^2-1)/(n^2-1)$.
\end{theorem}

Theorem~\ref{prop:nearorth} says that if $\bB$ and $\bE$ are nearly orthogonal Latin hypercubes,
the resulting Latin hypercube $\bL$ is also nearly orthogonal.
An example, illustrating the use of this result, is given below.

\begin{table}[h]
\caption{Design matrix of $\bB_0$ in Example~\ref{exam:nearorth}}
\label{Table:nolhd16}
\begin{center}
\scalebox{0.75}{
\begin{tabular}{c}
$\left(
\begin{array}{rrrrrrrrrrrrrrr}
 -15&  15& -13&  13&  -5& -13&   5&   3&  -1&   5 & -7 &  5&  -9 & -9&
5\\
 -13& -15&  -3&   3&   7&   3&  15& -11&  13&  -5 &  7 &-13&  -7 & -3&  -3  \\
 -11&  -9&  -5& -11& -15&  13&  -5&  11&  -9&   9 &  9 &  3&  -5 & -1& -11  \\
  -9&  -1&   9& -15& -11&   1&  -1& -13&   5&  -1 &-15 &  7&   1 &  3&  15  \\
  -7&   1&  -7&   7&  15&  15& -13&   9&  -5& -13 & -3 & -1&  -1 &  7&  13  \\
  -5&  13&  11&  -5&   9&  -7&  -3&  -9& -13&  11 & 13 & -9&  -3 & 13&   1  \\
  -3&  -5&  13&  15&  -9&  -9& -11&   1&   7&  -9 & 15 & 11&   9 &  1&  -1  \\
  -1& -11&   3&  -7&  11& -15&  13&  15&  -7&  -3 & -9 &  9&   7 &  9&  -5  \\
   1&   3&  -9&  -3&  -1&  -5& -15&  -1&  11&   3 &-11 &-15&  15 &  5& -15  \\
   3&  -3&  15&  11&   3&   9&   1&  -7& -15&   1 &-13 & -3&   3 &-15&  -9  \\
   5&   9&   7&  -1&   5&  11&   9&  13&  15&  15 &  5 &  1&  11 & -7&   9  \\
   7&   7&  -1& -13&  13&  -1&  -7&  -5&   9&  -7 &  3 & 15& -13 &-11& -13  \\
   9&   5& -11&  -9&  -7&  -3&   7&  -3& -11& -15 & 11 & -7&  13 &-13&   7  \\
  11&  11&   5&   5& -13&   7&  11&   5&   3& -11 & -5 & -5& -11 & 15&  -7  \\
  13&  -7& -15&   9&   1&   5&   3& -15&  -3&  13 &  1 & 13&   5 & 11&   3  \\
  15& -13&   1&   1&  -3& -11&  -9&   7&   1&   7 & -1 &-11& -15 & -5&  11
\end{array}\right)$
\end{tabular}}
\end{center}
\end{table}

\begin{example}\label{exam:nearorth}
Let $\bA=(1,1)^\prime$ and $\bE=(1/2,-1/2)^\prime$. Choose a $16
\times 15$ nearly orthogonal Latin hypercube $\bB=\bB_0/2$ where $\bB_0$ is displayed in Table~\ref{Table:nolhd16},
and $\bB$ has $\rho_{ave}^2(\bB)=0.0003$ and $\rho_M(\bB)=0.0765$. Taking any
15 columns of a Hadamard matrix of order 16 to be $\bF$ and then applying
(\ref{eq:kron}), we obtain a Latin hypercube $\bL$ of 32 runs
and 15 factors. As $\rho_{ave}^2(\bE)=\rho_M(\bE)=0$, we have
$\rho_{ave}^2(\bL)=(n_2^2-1)^2\rho_{ave}^2(\bB)/(n^2-1)^2=0.00002$
and $\rho_M(\bL)=(n_2^2-1)\rho_M(\bB)/(n^2-1)=0.0191$.
\end{example}

\subsubsection{Existence of orthogonal Latin hypercubes }\label{subsubsec:orth4}

A problem, of at least theoretical importance, in the study of orthogonal Latin hypercubes is to determine the maximum number $k^*$ of columns in an orthogonal Latin hypercube of a given run size $n$. Theorem~\ref{theo:exist} tells us that $k^* =1$  if $n$ is 3 or $n = 4m + 2$ for any non-negative integer $m$ and $k^* \geq 2$ otherwise. \cite{lin2010new} obtained a stronger result.
\begin{theorem}\label{prop:m}
The maximum number $k^*$ of factors for an orthogonal
Latin hypercube of  $n = 16m + j$ runs
has a lower bound given below: \\
(i) $k^* \geq 6$ for all $n=16m+j$ where $m \geq 1$ and
$j \neq 2, 6, 10, 14$; \\
(ii) $k^* \geq 7$ for $n=16m + 11$ where $m \geq 0$; \\
(iii) $k^* \geq 12$ for $n=16m, 16m + 1$ where $m \geq 2$; \\
(iv) $k^* \geq 24$ for $n=32m, 32m + 1$ where $m \geq 2$; \\
(v) $k^* \geq 48$ for $n=64m, 64m + 1$ where $m \geq 2$.
\end{theorem}

The above theorem provides a general lower bound on the maximum number $k^*$ of factors for an orthogonal Latin hypercube of $n$ runs.
We now summarize the results on the best lower bound  on  the maximum number $k^*$ in an $\OLH(n,k^*)$ from all existing approaches for $n \leq 256$.
Table~\ref{table:small} lists the best lower bound on the maximum number $k^*$ in an $\OLH(n,k^*)$ for $n \leq 24$. These values except the case $n=16$ were obtained by \cite{lin2008} through an algorithm. For $n=16$,  \cite{steinberg2006construction} obtained an orthogonal Latin hypercube with 12 columns. Table~\ref{table:large} reports the best lower bound on
the maximum number $k^*$ in an $\OLH(n,k^*)$ for $24 <  n \leq 256$ as well as the corresponding approach for achieving
the best lower bound.
\begin{table}[t]
\caption{The best lower bound $k$ on the maximum number $k^*$ of
factors in $\OLH(n,k^*)$ for $n \leq 24$ }
\begin{center}
\scalebox{0.9}{
\begin{tabular}{l|cccccccccccccccc}
\hline \hline
$n$ & 4 & 5 & 7 & 8 & 9 & 11 & 12 & 13 & 15 & 16 & 17 & 19 & 20 & 21 & 23 & 24 \\
\hline $k$ & 2 & 2 & 3 & 4 & 5 & 7 & 6 & 6 & 6 & 12 & 6 & 6 & 6& 6 & 6 & 6\\
\hline
\end{tabular}}
\end{center}\label{table:small}
\end{table}

\begin{table}[t]
\caption{The best lower bound $k$ on the maximum number $k^*$ of
factors in $\OLH(n,k^*)$ for $n>24$ }
\begin{center}
\scalebox{0.8}{
\begin{tabular}{l l lll ccccccc lllll}
\hline \hline
$n$&&$k$& &Reference             & & & & & & &  $n$&& $k$&& Reference    \\
\hline
25 & &12& &\cite{lin2009construction}& & & & & & &  144&& 24 && \cite{lin2010new} \\
32 & &16 & &\cite{sun2009construction} & & & & & & &  145&& 12 && \cite{lin2010new} \\
33 & &16 & &\cite{sun2009construction}& & & & & & &  160&& 24 && \cite{lin2010new} \\
48 & &12 & &\cite{lin2010new}     & & & & & & &  161&& 24 && \cite{lin2010new} \\
49 & &24 & &\cite{lin2009construction}       & & & & & & &  169&& 84 &&  \cite{lin2009construction} \\
64 & &32 & &\cite{sun2009construction} & & & & & & &  176&& 12 &&\cite{lin2010new} \\
65 & &32 & &\cite{sun2009construction} & & & & & & &  177&& 12 &&\cite{lin2010new} \\
80 & &12 & &\cite{lin2010new}    & & & & & & &  192&& 48 && \cite{lin2010new} \\
81 & &50 & &\cite{lin2009construction}       & & & & & & &  193&& 48 && \cite{lin2010new} \\
96 & &24 & &\cite{lin2010new}       & & & & & & &  208&& 12 && \cite{lin2010new}\\
97 & &24 & &\cite{lin2010new}    & & & & & & &  209&& 12 && \cite{lin2010new} \\
112& &12 & &\cite{lin2010new}      & & & & & & &  224&& 24 && \cite{lin2010new} \\
113& &12 & &\cite{lin2010new}     & & & & & & &  225&& 24 && \cite{lin2010new} \\
121& &84 & &\cite{lin2009construction}       & & & & & & &  240&& 12 && \cite{lin2010new} \\
128& &64 & &\cite{sun2009construction} & & & & & & &  241&& 12 && \cite{lin2010new}  \\
129& &64 & &\cite{sun2009construction} & & & & & & &  256&& 248&&\cite{steinberg2006construction}\\
\hline
\end{tabular}}
\end{center}\label{table:large}
\end{table}

\section{Other Space-filling Designs}\label{sec:other}



Section~\ref{sec:LHD} discussed various Latin hypercube designs that are suitable for computer experiments. A Latin hypercube design does not have repeated runs and each of its factors has as many levels as the run size.
\cite{bingham2009orthogonal}  argued that it is absolutely unnecessary to have the same number
of levels as the run size in many practical applications.
They proposed the use of orthogonal and nearly orthogonal designs   for computer experiments, where each factor is allowed to have repeated levels.  This is a rich class of orthogonal designs, including two-level orthogonal designs and orthogonal Latin hypercubes as special cases. This section reviews the concept and constructions of orthogonal designs. We also review another class of orthogonal designs provided by \cite{MoonDeanSantner2011}. Other classes of space-filling designs that do not fall under Latin hypercube designs are
low-discrepancy sequences and uniform designs. Both types of designs originate from the field of numerical analysis and give rise to attractive space-filling designs.
We provide a brief account of low-discrepancy sequences and review various measures of uniformity in uniform designs.

\subsection{Orthogonal designs with many levels}\label{subsec:other1}

Consider designs of $n$ runs for $k$ factors each of $s$ levels, where $2 \leq s \leq n$.
For convenience, the $s$ levels are chosen to be centered and equally spaced; one such choice is $-(s-1)/2, -(s-3)/2,\ldots, (s-3)/2, (s-1)/2$. Such a design is denoted by $\D(n,s^k)$ and can be represented by an $n \times k$ design matrix $\bD=(d_{ij})$ with entries from the above set of $s$ levels. A Latin hypercube of $n$ runs for $k$ factors is a $\D(n,s^k)$ with $n=s$.

Let $\bA=(a_{ij})$ be an $n_1 \times k_1$ matrix with entries $a_{ij} =\pm 1$ and $\bD_0$ be a $\D(n_2,s^{k_2})$. \cite{bingham2009orthogonal} constructed the $(n_1n_2)\times (k_1k_2)$ design
\begin{equation}\label{eq:bst}
\bD = \bA \otimes \bD_0.
\end{equation}

If $\bA$ is column-orthogonal, then design $\bD$ in (\ref{eq:bst}) is orthogonal if and only if $\bD_0$ is orthogonal.
 This provides a powerful way to construct a rich class of orthogonal designs for computer experiments, as illustrated by
 Example~\ref{exam:bst}.

\begin{example}\label{exam:bst}
Let $\bD_0$ be the orthogonal Latin hypercube $\OLH(16,12)$ constructed by \cite{steinberg2006construction}.
The construction method in (\ref{eq:bst}) gives a series of orthogonal designs of $16m$ runs for $12m$ factors by letting $\bA$ be a Hadamard matrix of order $m$, where $m$ is an integer such that a Hadamard matrix  of order $m$ exists.
\end{example}

Higher order orthogonality and near orthogonality of $\bD$ in (\ref{eq:bst}) were also discussed in \cite{bingham2009orthogonal}.
They considered two generalizations of the method (\ref{eq:bst}). Let $\bD_j$ be a $\D(n_2,s^{k_2})$, for each
$j=1,\ldots, k_1$. One generalization is
\begin{equation}\label{eq:bst2}
\bD=(a_{ij}\bD_j)=\left [
\begin{array}{cccc}
  a_{11}\bD_1   &  a_{12}\bD_2  &   \ldots  &   a_{1k_1}\bD_{k_1}  \\
  a_{21}\bD_1   &  a_{22}\bD_2   &   \ldots  &   a_{2k_1}\bD_{k_1}  \\
  \vdots    &  \vdots    &   \ddots  &   \vdots   \\
  a_{n_11}\bD_1 &  a_{n_12}\bD_2 &   \ldots  &   a_{n_1k_1}\bD_{k_1} \\
\end{array}\right].
\end{equation}

The following results study the orthogonality of design $\bD$ in (\ref{eq:bst2}).

\begin{theorem}\label{prop:gen}
Let $\bA$ be column-orthogonal. We have that\\
(i) $\rho_M(\bD) = \hbox{max} \{ \rho_M(\bD_1), \ldots, \rho_M(\bD_{k_1})\}$,\\
(ii) $\rho^2_{ave}(\bD) = w[\rho^2_{ave}(\bD_1) + \ldots +\rho^2_{ave}(\bD_{k_1}) ]/k_1$, where $w =(k_2-1)/(k_1k_2-1)$, and\\
(iii) $\bD$ in (\ref{eq:bst2}) is orthogonal if and only if $\bD_1,\ldots, \bD_{k_1}$ are all orthogonal.
\end{theorem}

The generalization (\ref{eq:bst2}) constructs designs with improved projection properties \cite[]{bingham2009orthogonal}.
The research on orthogonal designs was further pursued by  \cite{georgiou2010orthogonal} who proposed an alternative construction method and
obtained many new designs.

Another class of orthogonal designs is Gram-Schmidt designs constructed by \cite{MoonDeanSantner2011}. A Gram-Schmidt design for $n$ observations and $k$ inputs is generated from an $n \times k$ Latin hypercube design $\bD = (d_{ij}) =(\bd_1,\ldots,\bd_k)$ as follows.
\begin{myindentpar}{0.5cm}
\begin{itemize}
\item[Step 1:] Center the $j$th column of $\bD$ to have mean zero:
$$ \bv_j = \bd_j - \sum_{i=1}^n d_{ij}/n, \hbox{for} \ j=1,\ldots,k.$$
\item[Step 2:] Apply the Gram-Schimidt algorithm to $\bv_1,\ldots,\bv_k$ from Step~1 to form orthogonal columns
$$\bu_j = \left\{
  \begin{array}{ll}
    \bv_1, & j=1; \\
    \bv_j - \sum_{i=1}^{j-1} \frac{\bu_i\bv_j}{||\bu_i||^2}\bu_i, &  j=2,\ldots,k,
  \end{array}
\right.$$
\noindent where $||\bu_i||$ represents $l_2$ norm of $\bu_i$.
\item[Step 3:] Scale $\bu_j$ from Step 2 to the desired range and denote the resulting column by $\bx_j$. Set $\bX = (\bx_1,\ldots,\bx_k)$.
\end{itemize}
\end{myindentpar}
Any two columns of  design $\bX$ constructed via the three steps above have zero correlation.

\subsection{Low-discrepancy sequences and uniform designs}\label{subsec:other2}

Many problems in various fields such as quantum physics and computational finance require  calculating definite integrals of a function over a multi-dimensional unit cube. It is very common that the function may be so complicated  that the integral cannot be obtained analytically and precisely, which calls for numerical methods of approximating the integral.

Recall the numerical integration problem discussed in Section~\ref{subsec:intro}. The quantity $\hat{\mu}$ in (\ref{eq:muhat}) is used to approximate $\mu$ in (\ref{eq:mu}). Respecting the common notations, we use $s$ to denote the number of factors  and $\bchi = [0,1]^s$ the
design region  in this section.  Let $\mathcal{P}=(\bx_1,\ldots,\bx_n)$ be a set of $n$ points in $\bchi$.
The bound of the integration error is given by  Koksma-Hlawka inequality,
 \begin{equation}\label{eq:kh}
 \Big |\mu - \hat{\mu} \Big |  \leq V(f) D^*(\mathcal{P}),
 \end{equation}
\noindent where $V(f)$ is the variation of  $f$ in the sense of Hardy and Krause and $D^*(\mathcal{P})$ is the star discrepancy of   the $n$ points $\mathcal{P}$ \cite[]{weyl1916gleichverteilung} and described below. Motivated by the fact that the Koksma-Hlawka bound in (\ref{eq:kh}) is proportional to the star discrepancy of the points, different methods for generating points  in $\bchi$ with as small a star discrepancy as possible have been proposed. Such methods are referred to as quasi-Monte Carlo methods \cite[]{Niederreiter1992}.

For each $\bx=(x_1,\ldots,x_s)$ in $\bchi$, let $J_{\bx} = [0,\bx)$ denote the interval $[0,x_1) \times \cdots \times [0,x_s)$, $N(\mathcal{P}, J_{\bx})$ denote the number of points of $\mathcal{P}$ falling in $J_{\bx}$, and $\hbox{Vol}(J_{\bx} )$ be the volume of interval $J_{\bx} $. The {\em star discrepancy} $D^*(\mathcal{P})$ of $\mathcal{P}$ is defined by
\begin{equation}\label{eq:star}
D^*(\mathcal{P}) = \max_{\bx \in \bchi} \Big | \frac{
N(\mathcal{P}, J_{\bx})}{n} - \hbox{Vol}(J_{\bx}) \Big|.
\end{equation}
A sequence $S$ of points in $\bchi$ is called a {\em low-discrepancy sequence} if its first $n$ points have
$$D^*(\mathcal{P}) = O(n^{-1} (\hbox{log} \ n )^s),$$
where $O(\cdot)$ is big O notation. As a comparison, if the set $\mathcal{P}$ is chosen by the Monte Carlo method, that is, $\bx_1,\ldots, \bx_n$ are independent random samples from the uniform distribution, then
$D^*(\mathcal{P}) = O(n^{-1/2})$, which is considered too slow in many applications \cite[]{Niederreiter2012}.

Construction of low-discrepancy sequences is a very active research area in the study of quasi-Monte Carlo methods. There are many constructions available; examples are the good lattice point method, the good point method, Halton sequences, Faure sequences   and $(t,s)$-sequences.  For a comprehensive treatment of low-discrepancy sequences, see \cite{Niederreiter1992}.
Here we provide a brief account of two popular and most widely studied methods, {$(t,s)$-sequences} and   uniform designs.

\subsubsection{$(t,m,s)$-nets and $(t,s)$-sequences}\label{subsubsec:nets}

The definitions of  $(t,m,s)$-nets and $(t,s)$-sequences  require a concept of elementary intervals. An {\em elementary interval} in base $b$ is an interval $E$ in $[0,1)^s$ of the form
\begin{equation}\label{eq:eleint}
 E = \prod_{i=1}^s \Big[ \frac{a_i}{b^{d_i}}, \frac{a_i+1}{b^{d_i}} \Big)
 \end{equation}
\noindent with integers $a_i$ and $d_i$ satisfying $d_i \geq 0$ and $ 0 \leq a_i < b^{d_i}$.
For $i=1,\ldots,s$,  the $i$th axis of an elementary interval has length $b^{-d_i}$ and thus an elementary interval has a volume $b^{-\sum_{i=1}^s d_i}$.

For integers $b \geq 2$ and $0 \leq t \leq m$, a {\em $(t,m,s)$-net} in base $b$ is a set of $b^m$ points in $[0,1)^s$ such that every elementary interval in base $b$ of volume $b^{t-m}$ contains exactly $b^t$ points. For given values of $b$, $m$ and $s$, a smaller value of $t$ leads to a smaller elementary interval, and thus a set of points with better uniformity.  Consequently, a smaller value of $t$ in $(t,m,s)$-nets in base $b$ is preferred.

An infinite sequence of points $\{\bx_n\}$ in $[0,1)^s$ is a {\em $(t,s)$-sequence} in base $b$ if for all $k \geq 0$ and $m > t$, the finite sequence $\bx_{kb^m+1}, \ldots, \bx_{(k+1)b^m}$ forms a $(t,m,s)$-net in base $b$.  Example~\ref{ex:seq} illustrates both concepts.

\begin{example}\label{ex:seq}
Consider a $(0,2)$-sequence in base $2$. Its first 8 points form a $(0,3,2)$-net in base $2$ and are displayed in Figure~\ref{fig:sq1} with $t=0, m=3, s=2$.
There are four types of elementary intervals in base $2$ of volume $2^{-3}$ with $(d_1,d_2)$'s in (\ref{eq:eleint}) being $(0,3)$, $(3,0)$, $(1,2)$, and $(2,1)$.  Figures~\ref{fig:seq11} - \ref{fig:seq14}  show a $(0,3,2)$-net in base $2$ when elementary intervals are given by $(d_1,d_2)=(0,3)$,
$(d_1,d_2)=(3,0)$, $(d_1,d_2)=(1,2)$, and $(d_1,d_2)=(2,1)$, respectively.  Note that in every elementary interval of the form
$$ \Big[ \frac{a_1}{2^{d_1}},\frac{ (a_1+1)}{2^{d_1}} \Big) \times
 \Big[ \frac{a_2}{2^{d_2}},\frac{ (a_2+1)}{2^{d_2}} \Big), \ 0 \leq a_i <2^{d_i}, \ i=1,2,$$
\noindent there is exactly one point.   The next  8 points in this (0,2)-sequence in base 2 also form a $(0,3,2)$-net in base 2.
The totality of all 16 points is a $(0,4,2)$-net in base 2. Analogous to Figure~\ref{fig:sq1}, Figure~\ref{fig:sq2} exhibits the $(0,4,2)$-net in  base 2 when elementary intervals are given by all $(d_1,d_2)$'s that satisfy $d_1+d_2=m=4$.
\begin{figure}[htp]
\centering
\subfigure[$d_1=0,d_2=3$]{\label{fig:seq11} \includegraphics[width=0.25\textwidth]{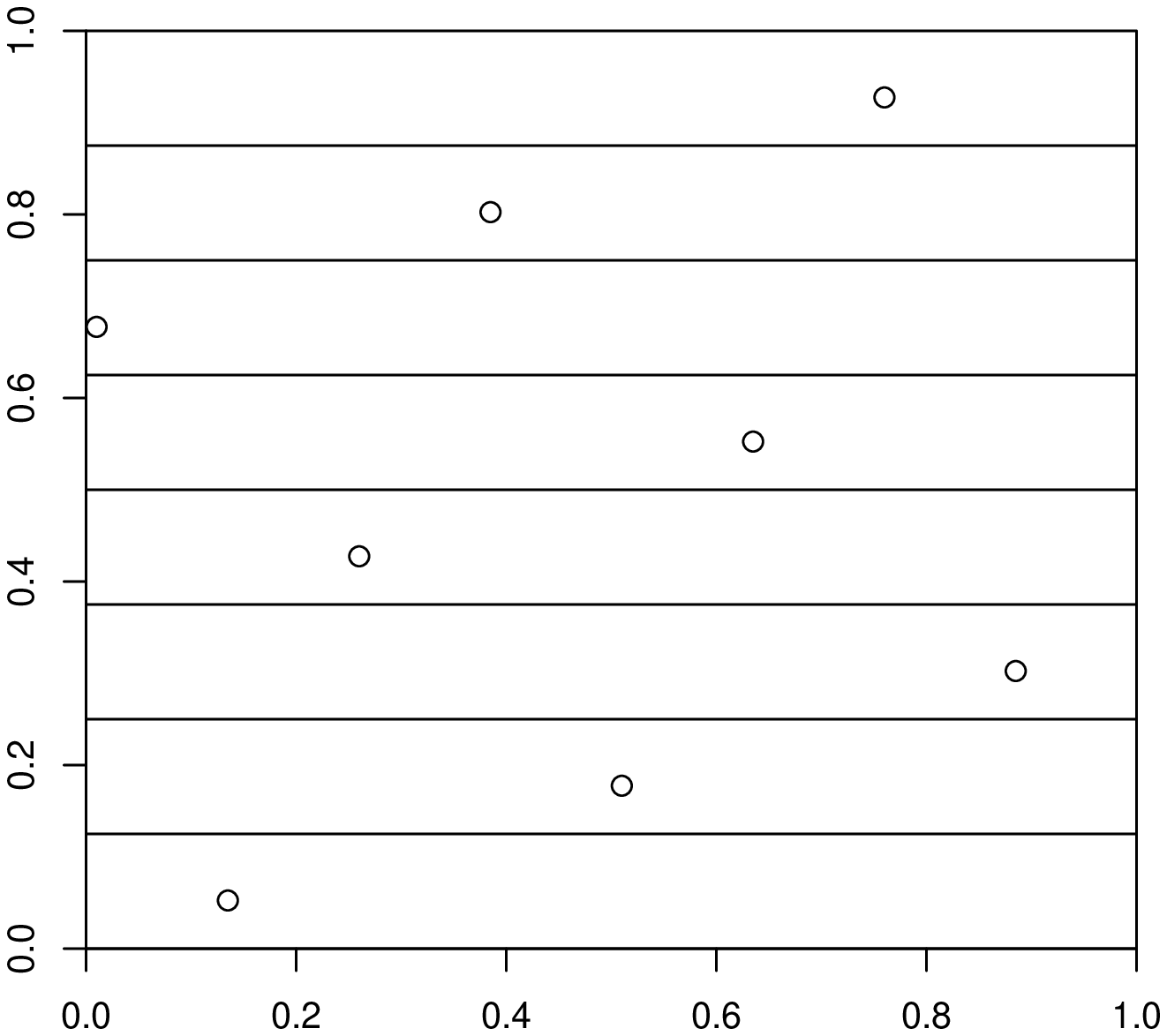} }\hspace{0.1in}
\subfigure[$d_1=3,d_2=0$]{\label{fig:seq12}  \includegraphics[width=0.25\textwidth]{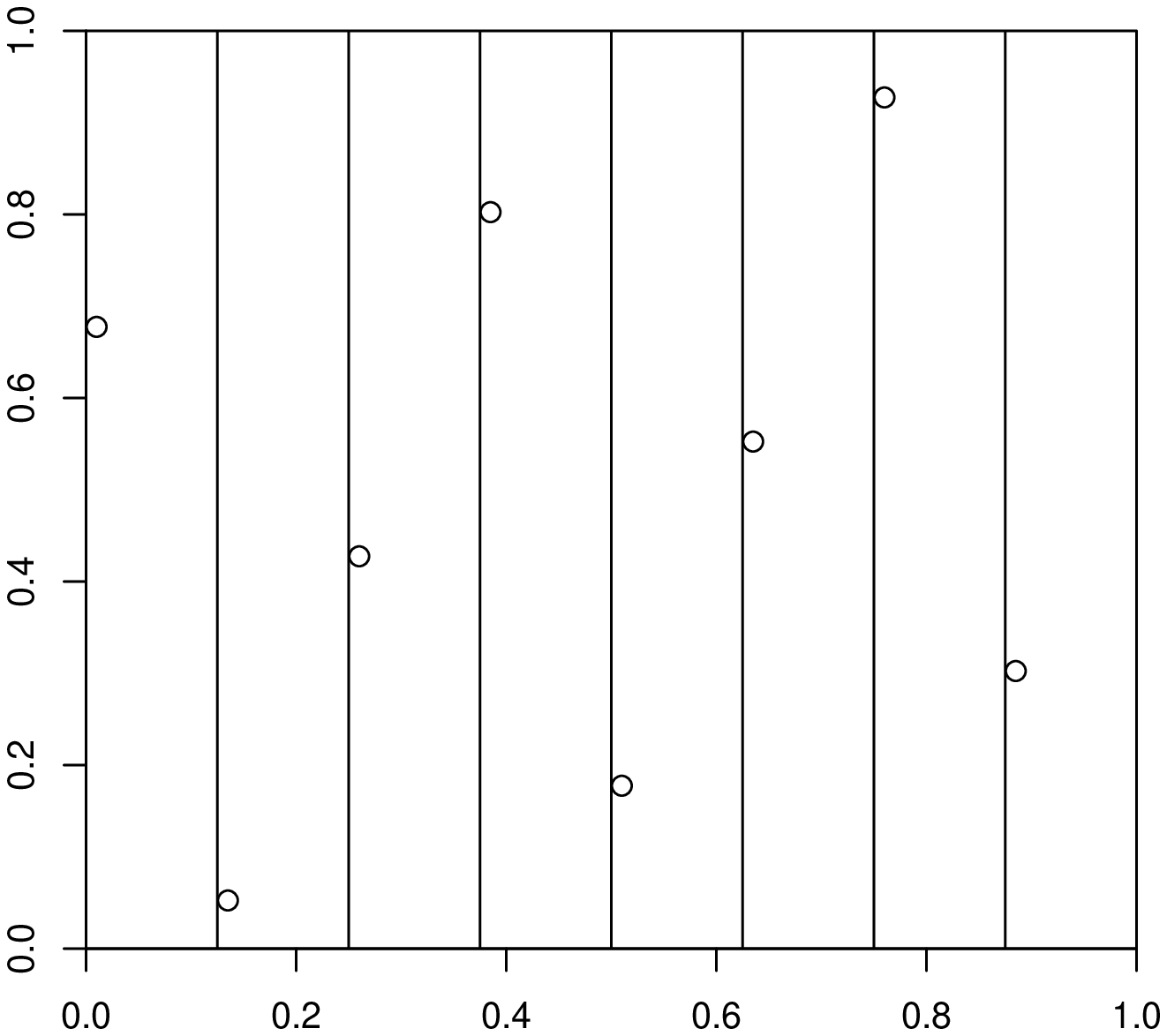}}\\
\subfigure[$d_1=1,d_2=2$]{\label{fig:seq13} \includegraphics[width=0.25\textwidth]{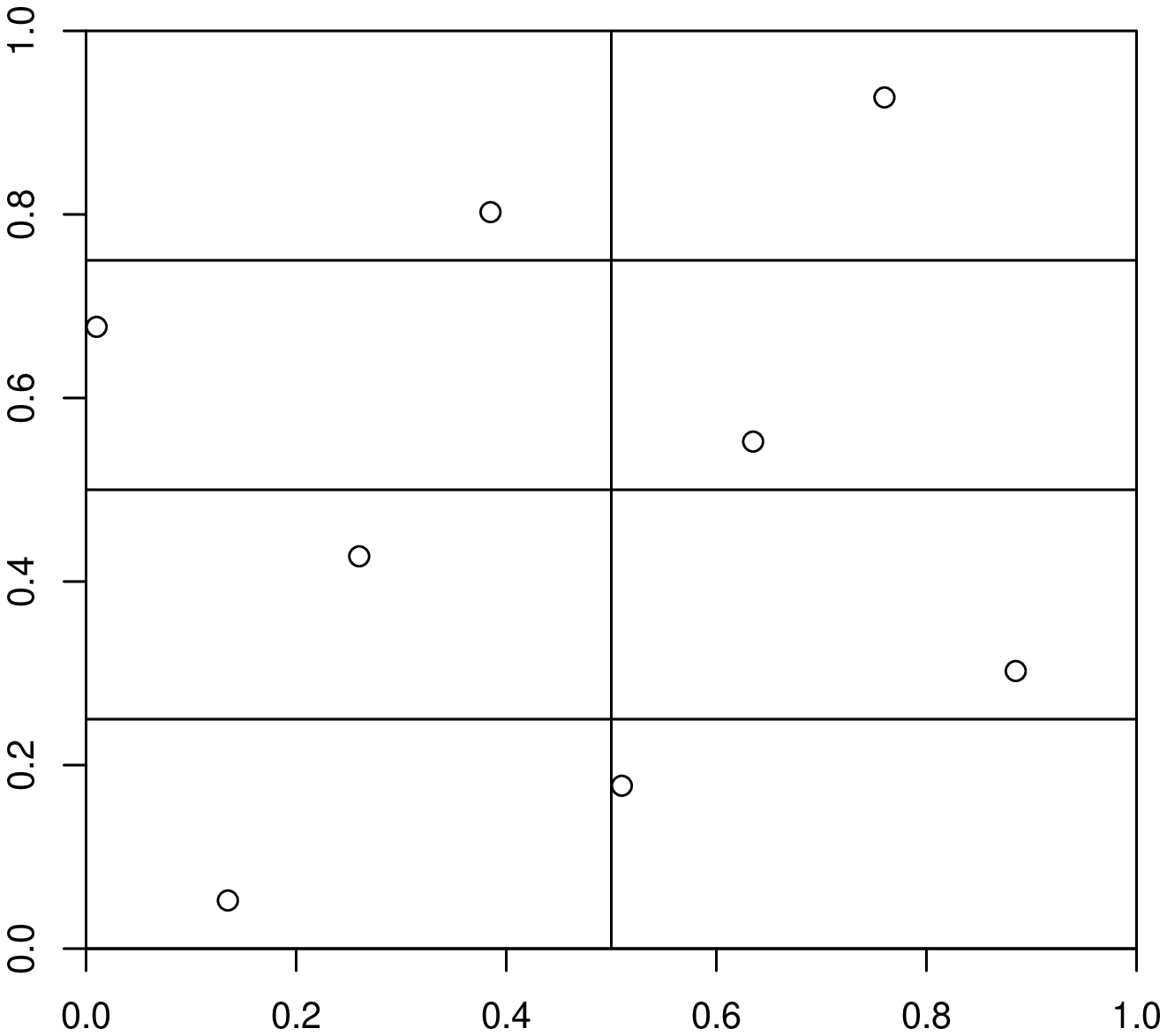} }\hspace{0.1in}
\subfigure[$d_1=2,d_2=1$]{\label{fig:seq14}  \includegraphics[width=0.25\textwidth]{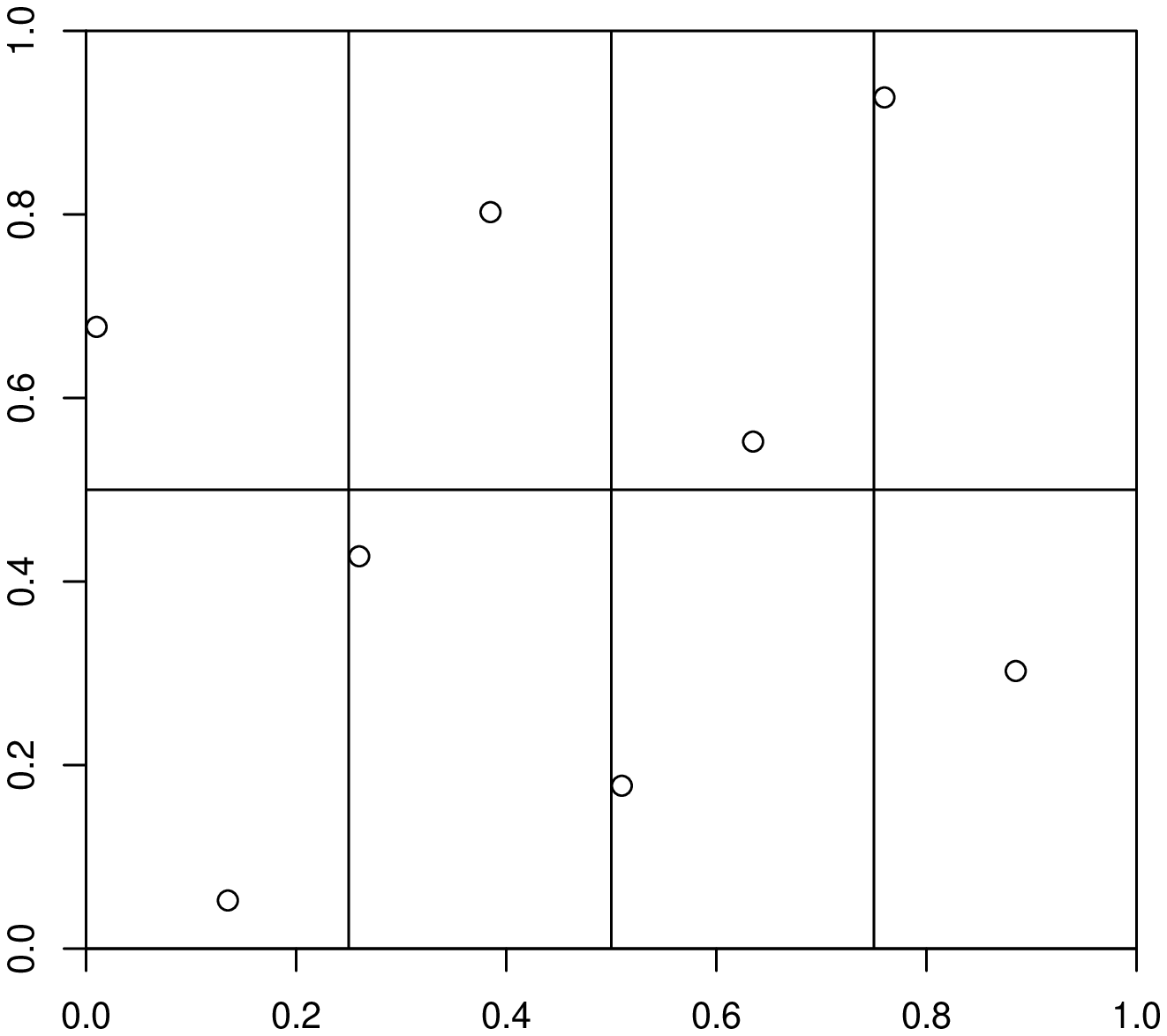}}
\caption{A $(0,3,2)$-net in base 2 seen using four types of elementary intervals }\label{fig:sq1}
\end{figure}

\begin{figure}[htp]
\centering
\subfigure[$d_1=2,d_2=2$]{\label{fig:seq21} \includegraphics[width=0.25\textwidth]{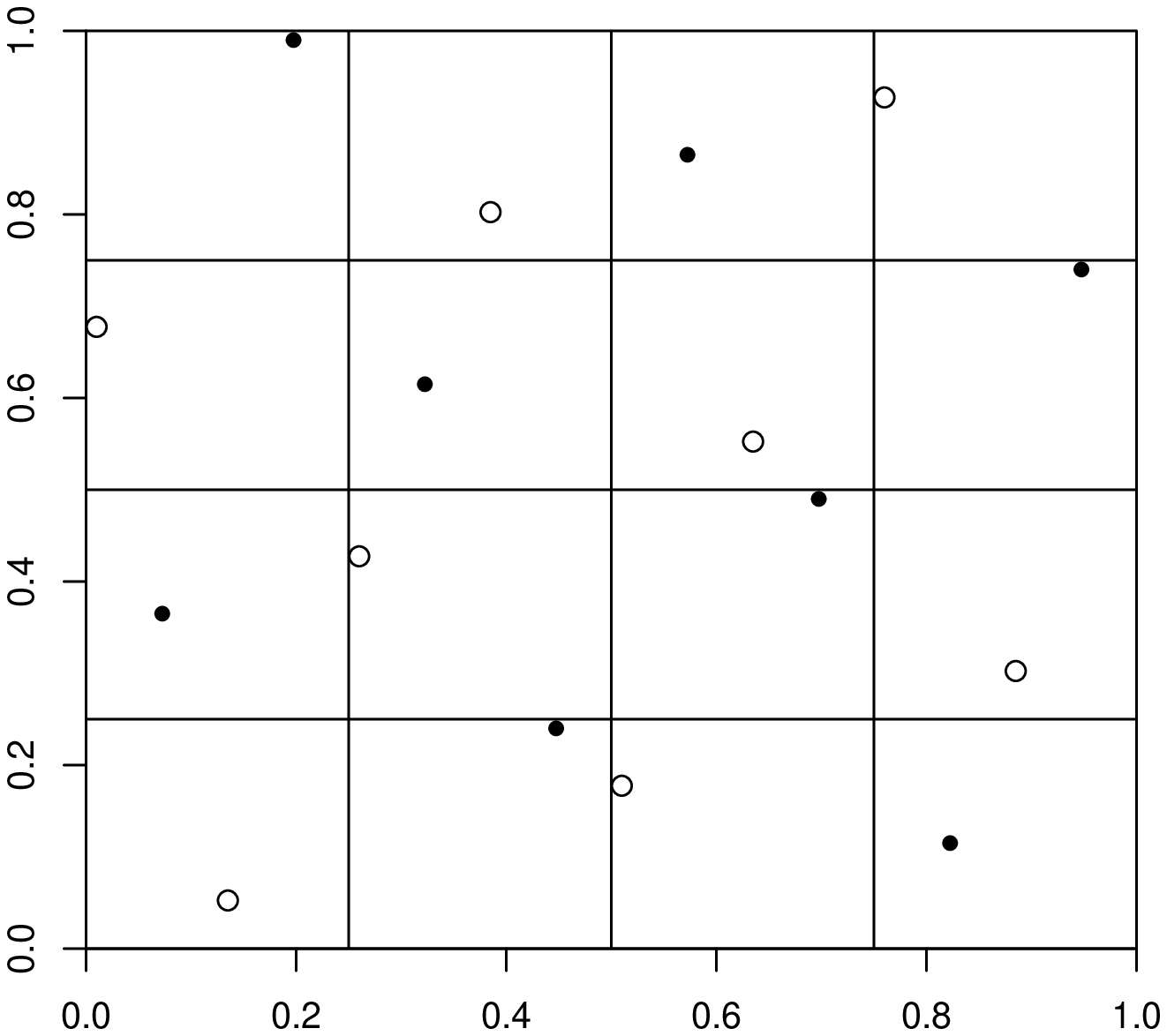} }\hspace{0.1in}
\subfigure[$d_1=3,d_2=1$]{\label{fig:seq22} \includegraphics[width=0.25\textwidth]{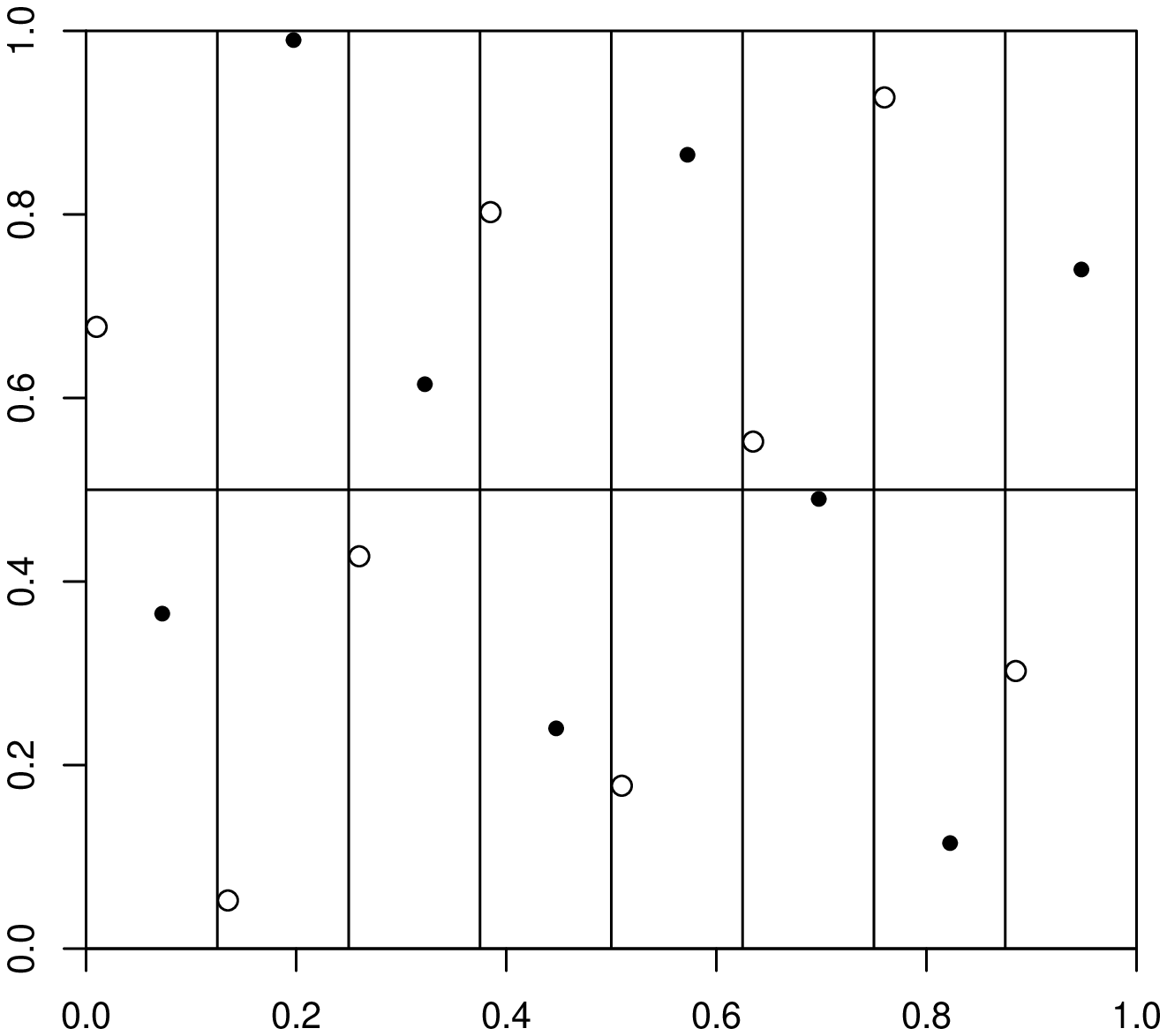} }\hspace{0.1in}
\subfigure[$d_1=1,d_2=3$]{\label{fig:seq23}  \includegraphics[width=0.25\textwidth]{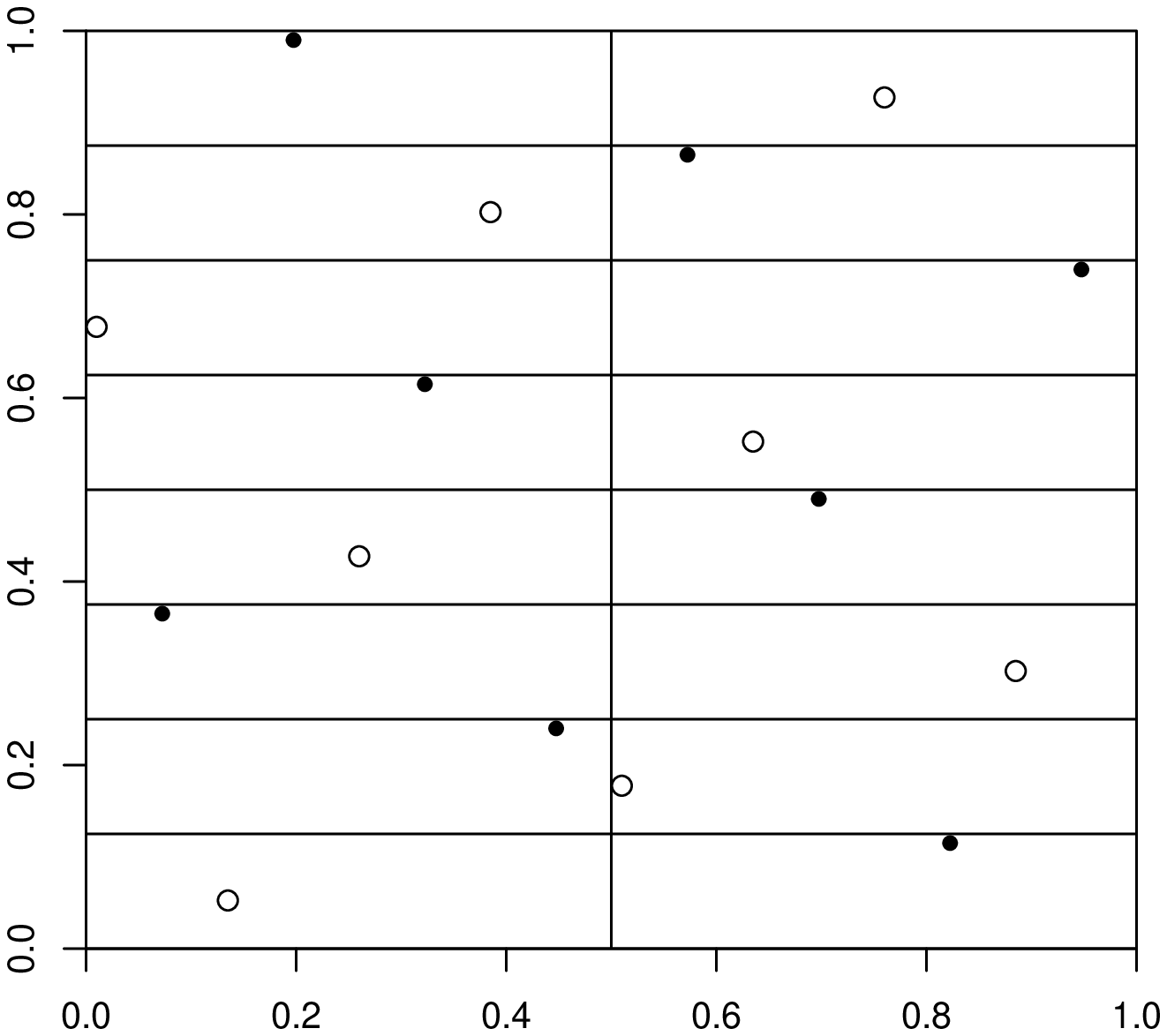}}\\
\subfigure[$d_1=4,d_2=0$]{\label{fig:seq24} \includegraphics[width=0.25\textwidth]{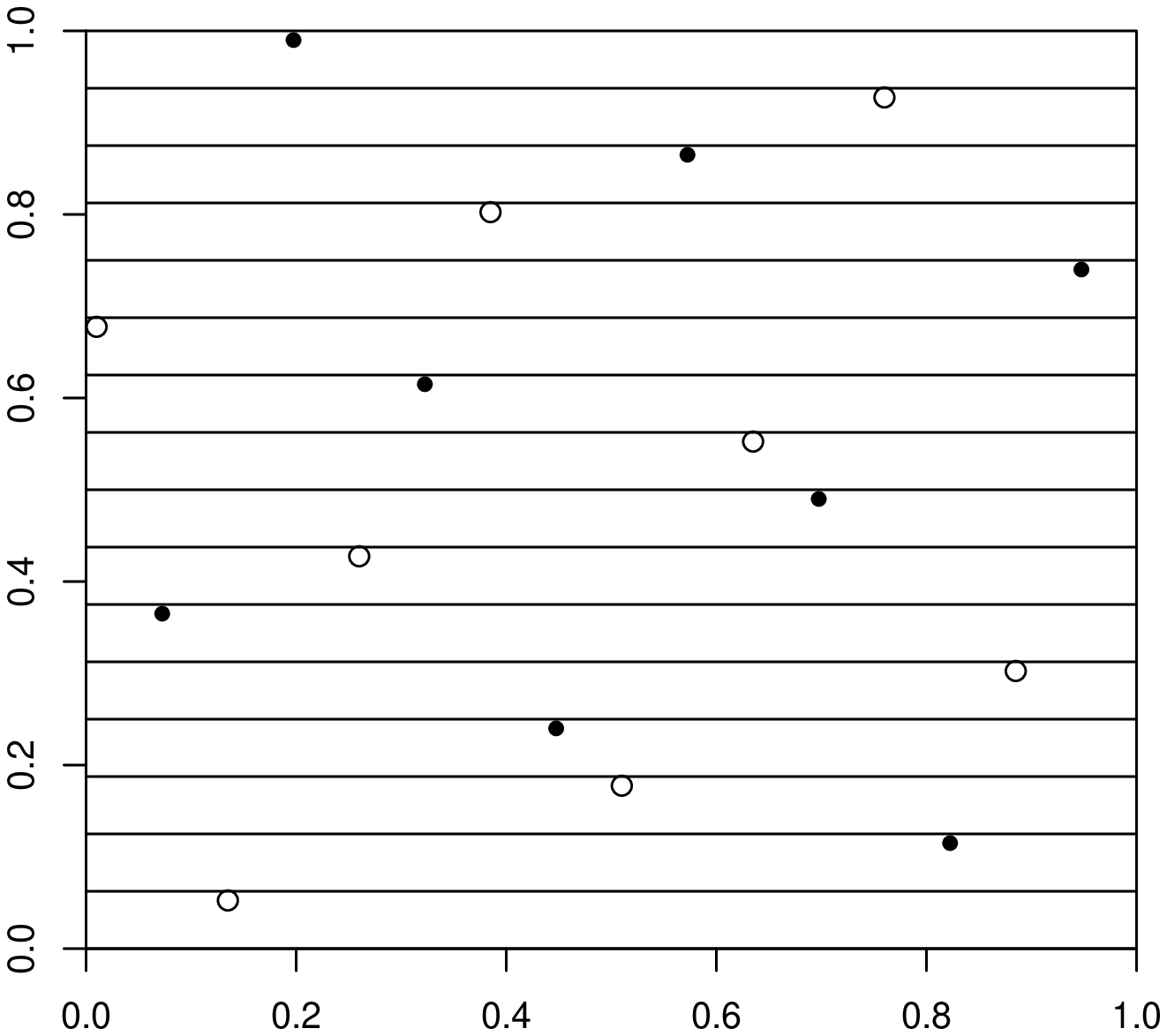} }\hspace{0.1in}
\subfigure[$d_1=0,d_2=4$]{\label{fig:seq25}  \includegraphics[width=0.25\textwidth]{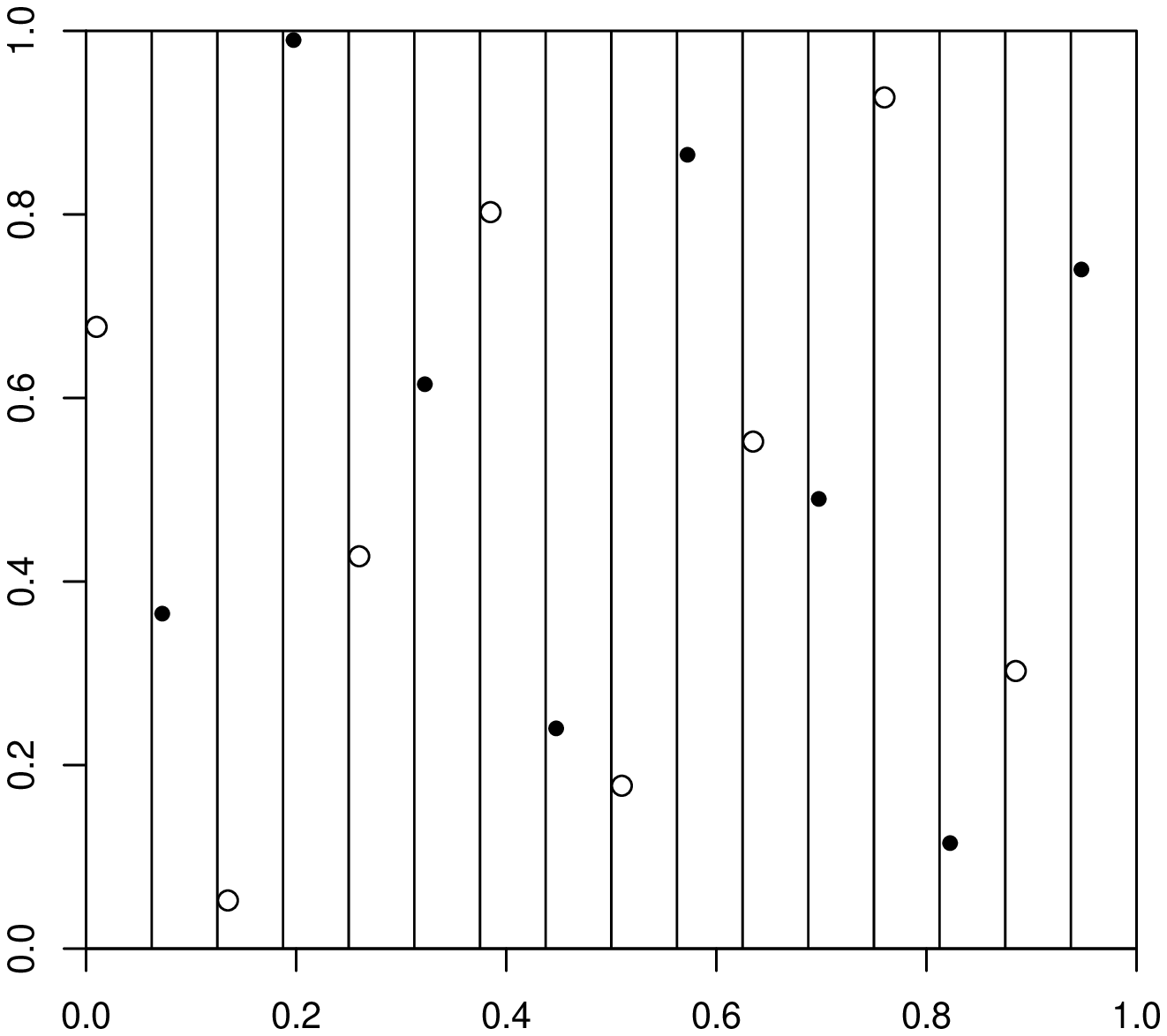}}
\caption{A $(0,4,2)$-net in base 2 seen using five types of elementary intervals,
the first and second 8 points are represented by $\circ$ and $\bullet$ }\label{fig:sq2}
\end{figure}
\end{example}


A general theory of $(t,m,s)$-nets and $(t,s)$-sequences was developed by
\cite{niederreiter1987point}. Some special cases of $(t,s)$-sequences are as follows.
Sobol' sequences \cite[]{sobol1967distribution} are $(t,s)$-sequences in base 2.
Faure sequences \cite[]{faure1982} are $(0,s)$-sequences in base $q$ where $q$ is a prime with $s \leq q$.  Niederreiter sequences \cite[]{niederreiter1987point} are $(0,s)$ sequences in base $q$ where $q$ is a prime or a prime power with $ s \leq q$.
Niederreiter-Xing sequences \cite[]{niederreiter1996low}  are $(t,s)$-sequences
in base $q$  for some certain $t$ where $q$ is a prime or a prime power with $s>q$.
For constructions of all these sequences, we refer the readers to \cite{niederreiter2008nets}.
Results on existing $(t,s)$-sequences are available in \cite{schurer2010mint}.

\subsubsection{Uniform designs}\label{subsubsec:uniform}
Motivated by the Koksma-Hlawka inequality in (\ref{eq:kh}), Fang  and Wang \cite[]{fang1980uniform,wang1981note}  introduced uniform designs, and by their definition, a uniform design is  a set of design points with the smallest discrepancy among all possible designs of the same run size. One choice of discrepancy is the star discrepancy in (\ref{eq:star}). More generally, one can use the {\em $L_p$ discrepancy},
\begin{equation*}\label{eq:lp}
D_p(\mathcal{P}) = \Big [ \int_{\bchi} \big| \frac{N(\mathcal{P}, J_{\bx} )}{n} - \hbox{Vol} (J_{\bx} )\big|^p d \bx\Big]^{1/p},
\end{equation*}
\noindent where $N(\mathcal{P}, J_{\bx} )$  and  $\hbox{Vol} (J_{\bx} )$ are defined as in (\ref{eq:star}).  Two special cases of the $L_p$ discrepancy are the $L_{\infty}$ discrepancy, which is the star discrepancy, and the $L_2$ discrepancy. While the $L_{\infty}$ discrepancy is difficult to compute, the $L_2$ discrepancy is much easier to evaluate because of a simple formula given by \cite{warnock},
$$D_2(\mathcal{P}) = 2^{-s} - \frac{2^{1-s}}{n} \sum_{i=1}^n \prod_{l=1}^s (1-x_{il}^2) + \frac{1}{n^2} \sum_{i=1}^n \sum_{j=1}^n \prod_{l=1}^s [1-\hbox{max}(x_{il}, x_{jl}) ],$$
\noindent where $x_{il}$ is the setting of the $l$th factor in the $i$th run, $i=1,\ldots, n$ and $l=1,\ldots,s$.

The $L_p$ discrepancy aims to achieve uniformity in the $s$-dimensional design space. Designs with the smallest $L_p$ discrepancy do not necessarily perform well in terms of projection uniformity in low dimensions.
\cite{hickernell1998generalized} proposed three new measures of uniformity, the symmetric $L_2$ discrepancy ($SL_2$), the centered $L_2$ discrepancy ($CL_2$), and the modified $L_2$ discrepancy ($ML_2$).
They are all defined through
\begin{equation}\label{eq:dmod}
D_{mod}(\mathcal{P}) = \sum_{u \neq \emptyset} \int_{\bchi_u} \Big | \frac{N(\mathcal{P}_{u},
J_{\bx_u})}{n}- \hbox{Vol}(J_{\bx_u}) \Big|^2 d u ,
\end{equation}
\noindent where $\emptyset$ represents the empty set, $u$ is a non-empty subset of the set $\{1,\ldots,s\}$, $|u|$ denotes the cardinality
of $u$, $\bchi_u$ is the $|u|$-dimensional unit cube involving the coordinates in $u$, $\mathcal{P}_u$
is the projection of the set of points $\mathcal{P}$ on $\bchi_u$,  $J_{\bx_u}$ is the projection of $J_{\bx}$ on $\bchi_u$, $N(\mathcal{P}_{u}, J_{\bx_u})$ denotes the number of points of $\mathcal{P}_{u}$ falling in $J_{\bx_u}$, and $\hbox{Vol}(J_{\bx_u})$ represents the volume of $J_{\bx_u}$. The symmetric $L_2$ discrepancy
chooses $J_{\bx}$ such that it is invariant if $x_{il}$ is replaced by $1-x_{il}$, $i=1,\ldots, n$ and $l=1,\ldots,s$, and it has the formula
$$(SL_2(\mathcal{P}))^2 = (\frac{4}{3})^s - \frac{2}{n} \sum_{i=1}^n \prod_{l=1}^s (1+2x_{il} - 2x_{il}^2)
+ \frac{2^s}{n^2} \sum_{i=1}^n \sum_{j=1}^n \prod_{l=1}^s \big( 1-|x_{il}-x_{jl}|\big).$$
The centered $L_2$ discrepancy chooses $J_{\bx}$ such that it is invariant under the reflections of $\mathcal{P}$ around any hyperplane with the $l$th coordinate being 0.5. Let $\mathcal{A}^s$ denote the set of $2^s$ vertices of the unit cube $\bchi$ and $\ba=(a_1,\ldots,a_s)\in \mathcal{A}^s$ be the closest one to $\bx$. The centered $L_2$ discrepancy takes $J_{\bx}$ in (\ref{eq:dmod}) to be
$$\{\bd=(d_1,\ldots,d_s) \in \bchi \ | \ \hbox{min}(a_j,x_j) \leq d_j < \hbox{max}(a_j,x_j), j=1,\ldots,s\}.$$
\noindent The formula for the centered $L_2$ discrepancy is given by
\begin{eqnarray*}
(CL_2(\mathcal{P}))^2 & = & (\frac{13}{12})^2 - \frac{2}{n} \sum_{i=1}^n \prod_{l=1}^s \big(
1+ \frac{1}{2}|x_{il}-0.5| - \frac{1}{2}|x_{il}-0.5|^2 \big) \\
&&+
\frac{1}{n^2} \sum_{i=1}^n\sum_{j=1}^n \prod_{l=1}^s \big( 1+\frac{1}{2}|x_{il}-0.5| +
\frac{1}{2}|x_{jl}-0.5| - \frac{1}{2} |x_{il}-x_{jl}| \big).
\end{eqnarray*}
The modified $L_2$ discrepancy takes $J_{\bx}=[\bzero,\bx)$ and has the formula
$$(ML_2(\mathcal{P}))^2  = (\frac{4}{3})^s - \frac{2^{1-s}}{n} \sum_{i=1}^n \prod_{l=1}^s
(3-x_{il}^2) + \frac{1}{n^2} \sum_{i=1}^n \sum_{j=1}^s \prod_{l=1}^s \big[
2-\hbox{max}(x_{il},x_{il}) \big].$$
For other discrepancy measures such as the wrap-around discrepancy, see \cite{fang2006design}.

Finding uniform designs based on a discrepancy criterion is an optimization problem.
However, searching uniform designs  in the entire unit cube  is computationally prohibitive for  large designs. Instead,
 it is convenient to find uniform designs within a class of  \emph{U-type designs}. Suppose that
each of the $s$ factors in an experiment has $q$ levels, $\{1,\ldots,q\}$. A U-type design, denoted by $U(n;q^s)$, is an $n \times s$ matrix in which the $q$ levels in each column appear equally often. Table~\ref{table:ud} displays a $U(6;3^2)$ and a $U(6;6^2)$.  For $q=n$, uniform $U$-type designs can be constructed by several methods such
as the good lattice method, the Latin square method, the expanding orthogonal array method and
the cutting method \cite[]{fang2006design}. For general values of $q$,  optimization algorithms have been considered, such as simulated annealing, genetic algorithm,  and threshold accepting \cite[]{Bohachevsky,Holland,WinkerFang1997}.
For more detailed discussions on the  theory and applications of uniform designs, see \cite{fang2000uniform} and  \cite{fang2003uniform}.
\begin{table}[!hbt]
\caption{Uniform designs $U(6;3^2)$ and $U(6;6^2)$}
\begin{center}
\begin{tabular}{ccc}
$U(6;3^2)$ & $\quad \quad \quad$ &  $U(6;6^2)$\\
$\begin{array}{rr}
1 & 1\\
2 & 2 \\
3 & 3\\
1 &3 \\
2 & 1\\
3 &2 \\
\end{array} $&&
$\begin{array}{rr}
1 & 3\\
2 & 5 \\
3 & 1\\
4 & 6\\
5 & 2\\
6 & 4\\
\end{array} $
\end{tabular}
\end{center}\label{table:ud}
\end{table}

\section{Concluding Remarks}\label{sec:con}

We have  provided an expository account of the constructions and properties of space-filling designs for computer experiments.
Research in this area remains active and will continue to thrive. Recently, a number of new directions have been pursued.   \cite{HeTang} introduced strong orthogonal arrays and associated Latin hypercubes,
while \cite{TangXuLin2012}  studied uniform fractional factorial designs.
Research has also been conducted to take advantage of many available results from other design areas such as factorial design theory, one such work being multi-layer designs proposed by \cite{BaJoseph}.
Another important direction is to develop methodology for the design regions in which input variables have dependency or constraints;
see \cite{Draguljic} and  \cite{Bowman} for more details.

\end{doublespace}
\begin{center}
\bibliographystyle{asa}
\bibliography{Lin-Tang}
\end{center}
\end{document}